\documentclass[sigconf]{acmart}
\AtBeginDocument{%
  \providecommand\BibTeX{{%
    \normalfont B\kern-0.5em{\scshape i\kern-0.25em b}\kern-0.8em\TeX}}}

\copyrightyear{2024}
\acmYear{2024}
\setcopyright{acmlicensed}\acmConference[SIGSPATIAL '24]{The 32nd ACM International Conference on Advances in Geographic Information Systems}{October 29-November 1, 2024}{Atlanta, GA, USA}
\acmBooktitle{The 32nd ACM International Conference on Advances in Geographic Information Systems (SIGSPATIAL '24), October 29-November 1, 2024, Atlanta, GA, USA}
\acmDOI{10.1145/3678717.3691304}
\acmISBN{979-8-4007-1107-7/24/10}

\usepackage{amssymb}
\usepackage{amsthm}
\usepackage{mathtools}
\usepackage{caption}
\usepackage{subcaption}
\usepackage{enumitem}
\usepackage{algorithm}
\usepackage{multicol}
\usepackage{multirow}
\usepackage{xcolor}

\usepackage[noend]{algpseudocode}
\usepackage{amsmath}
\algnewcommand\algorithmicforeach{\textbf{for each:}}
\algnewcommand\ForEach{\item[ \algorithmicforeach]}
\algdef{S}[FOR]{ForEach}[1]{\algorithmicforeach\ #1\ \algorithmicdo}
\algtext*{EndWhile}% Remove "end while" text
\algtext*{EndIf}% Remove "end if" text
\newcommand{\vx}{\mathbf{x}}
\newcommand{\vy}{\mathbf{y}}
\usepackage{amsmath}
\usepackage{xcolor}
\usepackage{bm}
\usepackage{hyperref}
\newcommand{\R}{\mathbb{R}}

\usepackage{algorithm}
\usepackage{algpseudocode}
\usepackage{amsmath}
\usepackage{bm}
\newcommand{\vz}{\mathbf{z}} % Assuming \vz is a vector like \vx and \vy
\newcommand{\veps}{\bm{\epsilon}}

\newcommand{\modelname}[1]{#1}
\newcommand{\grad}{\nabla}
\newcommand{\bepsilon}{\bm{\epsilon}}

\newcommand{\bx}{\bm{x}} % Defines \bx as bold x, modify if \bx_0 needs different formatting
\newcommand{\xeps}{\bm{\epsilon}}
\usepackage{amsmath}  % Essential for advanced math formatting
\usepackage{amssymb}  % Provides \mathbb and other math symbols
\usepackage{hyperref} % For handling hyperlinks and cross-referencing
\usepackage{cleveref} % Enhanced cross-referencing capabilities
\usepackage{bm}  % For bold math
\usepackage{titlesec}
\titlespacing*{\subsection}{0pt}{0.1\baselineskip}{0.2\baselineskip}

\AtBeginDocument{%
  \providecommand\BibTeX{{%
    \normalfont B\kern-0.5em{\scshape i\kern-0.25em b}\kern-0.8em\TeX}}}

\newtheorem{definition}{Definition}[section]
\begin{document}


\title{Towards Kriging-informed Conditional Diffusion for Regional Sea-Level Data Downscaling}

\author{Subhankar Ghosh}
\authornote{Subhankar Ghosh and Arun Sharma contributed equally to this paper.}
\email{ghosh117@umn.edu}
\affiliation{%
  \institution{University of Minnesota, Twin Cities}
  \city{Minneapolis}
  \state{Minnesota}
  \country{USA}
  \postcode{43017-6221}
}

\author{Arun Sharma}
\authornotemark[1]
\email{{sharm485@umn.edu}}
\affiliation{%
  \institution{University of Minnesota, Twin Cities}
  \city{Minneapolis}
  \state{Minnesota}
  \country{USA}
  \postcode{43017-6221}
}

% \author{Jayant Gupta}
% \email{jayant.j.gupta@oracle.com}
% \affiliation{%
%     \institution{Oracle Inc.}
%     \country{USA}
% }

\author{Jayant Gupta}
\email{jayant.j.gupta@oracle.com}
\affiliation{%
  \institution{Oracle Inc.}
  % \city{Minneapolis}
  % \state{Minnesota}
  \country{USA}
}

\author{Aneesh Subramanian}
\email{aneeshcs@colorado.edu}
\affiliation{%
  \institution{University of Colorado, Boulder}
  \city{Boulder}
  \state{Colorado}
  \country{USA}
  \postcode{43017-6221}
}

\author{Shashi Shekhar}
\email{shekhar@umn.edu}
\affiliation{%
  \institution{University of Minnesota, Twin Cities}
  \city{Minneapolis}
  \state{Minnesota}
  \country{USA}
  \postcode{43017-6221}
}
\renewcommand{\shortauthors}{Subhankar Ghosh, Arun Sharma, Jayant Gupta, Aneesh Subramanian, and Shashi Shekhar}

\begin{abstract}
Given coarser-resolution projections from global climate models or satellite data, the downscaling problem aims to estimate finer-resolution regional climate data, capturing fine-scale spatial patterns and variability. Downscaling is any method to derive high-resolution data from low-resolution variables, often to provide more detailed and local predictions and analyses. This problem is societally crucial for effective adaptation, mitigation, and resilience against significant risks from climate change. The challenge arises from spatial heterogeneity and the need to recover finer-scale features while ensuring model generalization. Most downscaling methods \cite{Li2020} fail to capture the spatial dependencies at finer scales and underperform on real-world climate datasets, such as sea-level rise. We propose a novel Kriging-informed Conditional Diffusion Probabilistic Model (Ki-CDPM) to capture spatial variability while preserving fine-scale features. Experimental results on climate data show that our proposed method is more accurate than state-of-the-art downscaling techniques.
\end{abstract}

%% \textcolor{red}{State-of-the-art downscaling techniques (Give examples)\cite{Li2020}}
%%
\begin{CCSXML}
<ccs2012>
   <concept>
       <concept_id>10010147.10010178.10010187.10010197</concept_id>
       <concept_desc>Computing methodologies~Spatial and physical reasoning</concept_desc>
       <concept_significance>500</concept_significance>
       </concept>
   <concept>
       <concept_id>10010147.10010178.10010224.10010240.10010243</concept_id>
       <concept_desc>Computing methodologies~Appearance and texture representations</concept_desc>
       <concept_significance>300</concept_significance>
       </concept>
   <concept>
       <concept_id>10010147.10010257.10010293</concept_id>
       <concept_desc>Computing methodologies~Machine learning approaches</concept_desc>
       <concept_significance>300</concept_significance>
       </concept>
 </ccs2012>
\end{CCSXML}

\ccsdesc[500]{Computing methodologies~Spatial and physical reasoning}
\ccsdesc[300]{Computing methodologies~Appearance and texture representations}
\ccsdesc[300]{Computing methodologies~Machine learning approaches}

\keywords{Remote Sensing, GeoAI, Climate Science, Geostatistics, Generative AI, Downscaling, Diffusion Models, Kriging}

\maketitle

\section{Introduction}
\begin{figure}[h]
    \centering
    \includegraphics[width=\columnwidth]{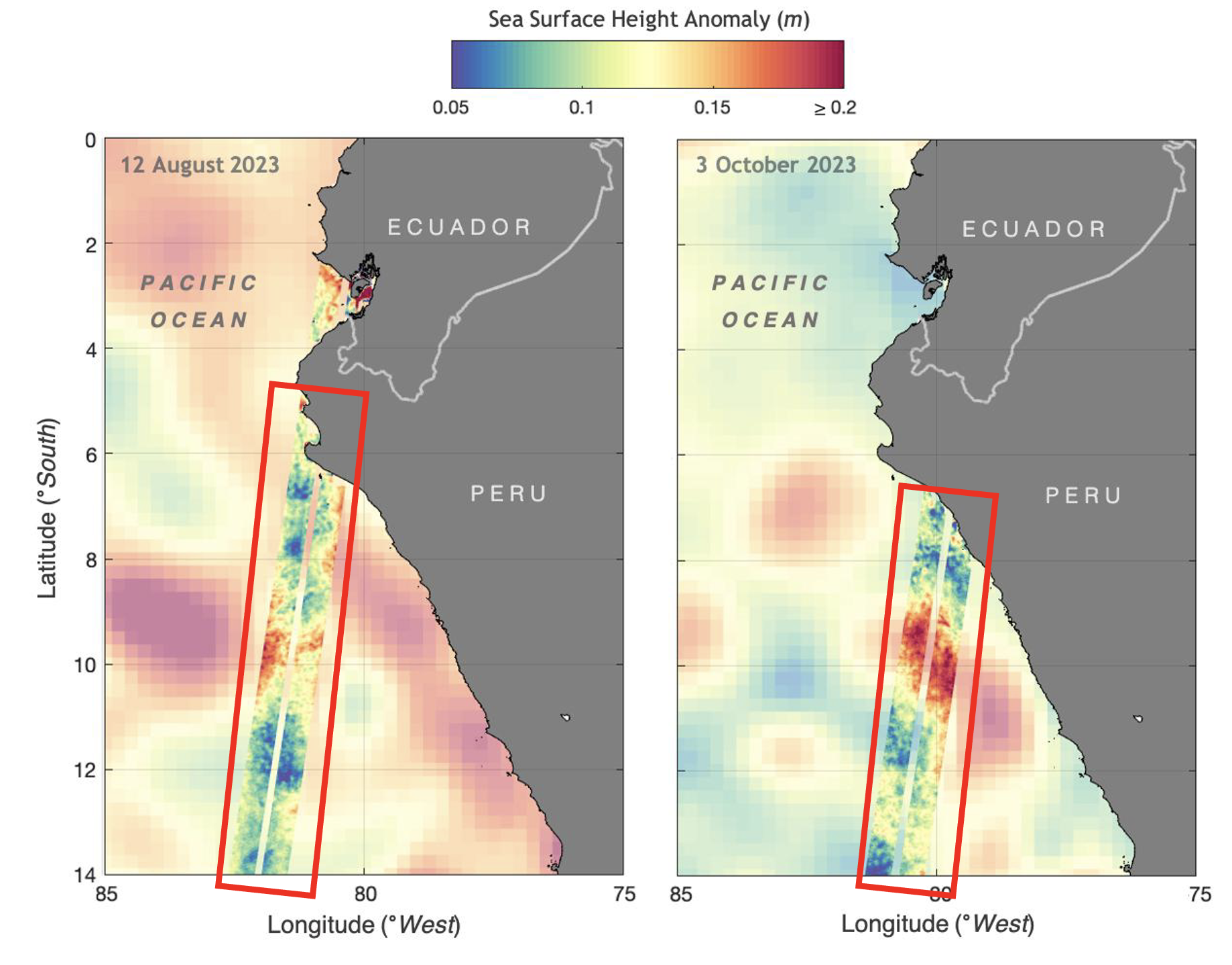}
    \caption{An illustrative example of differentiating coarse resolution and fine-scale resolution observations \cite{rutti2021entrepreneurship}. (Best in color)}
    \Description{An illustration showing the difference between coarse resolution and fine-scale resolution observations.}
    \label{fig:Problem_Statement}
\end{figure}

Given coarse-resolution climate projections from global climate models or satellite data, the problem of statistical downscaling aims to estimate high-resolution regional climate data, capturing fine-scale spatial patterns and variability for a climate variable (e.g., sea-level rise). Since downscaling generates high-resolution data from low-resolution variables, statistical downscaling uses statistical methods to establish relationships between coarse-resolution climate data and high-resolution historical observations. For instance, Figure \ref{fig:Problem_Statement} shows a geographic area near Ecuador and Peru in two different time frames. In both images, the blurry coarse-resolution projections display limited variation in sea surface height anomalies compared to the fine-scale resolution, as shown in the rectangular strips (in red). The ability to capture localized variations is crucial for accurately predicting climate change effects (e.g., sea-level rise) in a specific area.

The regional downscaling problem is vital for developing comprehensive climate policies and strategies. The impacts of climate change, as predicted by global climate models, range from extreme weather events to rises in sea levels and shifts in agricultural productivity. Enhancing the spatial precision of climate data is vital for accurately assessing risks in specific regions. Decision-makers require such information to strategize how communities can successfully mitigate and adapt to these threats. For example, coastal areas worldwide face severe risks associated with sea-level rise, where even small changes can lead to significant flooding, erosion, and habitat loss \cite{IPCC2021, Rabinowitz2023}. High-resolution data contributes to accurate regional predictions of this risk, essential for effective coastal management, infrastructure planning, and disaster preparedness.

% The problem of statistical downscaling in climate is crucial because it enables the translation of coarse-resolution climate data into high-resolution, localized climate projections. This enhanced spatial precision is vital for accurately assessing regional climate impacts, informing mitigation and adaptation strategies, and aiding decision-makers in managing risks associated with climate change, such as extreme weather events, sea level rise, and shifts in agricultural productivity. Sea-level rise is a significant consequence of climate change, posing severe risks to coastal regions worldwide. Hence, accurate predictions of sea-level changes are essential for effective coastal management, infrastructure planning, and disaster preparedness. Therefore, high-resolution sea-level data are vital for understanding the impacts of climate change on coastal areas, where even small changes can lead to significant consequences, such as increased flooding, erosion, and habitat loss. Furthermore, high-resolution data contribute to more reliable regional climate models for developing comprehensive policies and strategies.

% Accurate downscaled data enable more precise risk assessments and better-informed decision-making for adaptation and mitigation strategies. For instance, detailed sea-level projections are crucial for designing resilient infrastructure, implementing effective flood defenses, and planning evacuation routes in vulnerable coastal communities.

\begin{table}[]
\small
\centering
\caption{Applications of the Data Downscaling Problem}
\vspace*{-1\abovedisplayskip}
\label{tab:applications}
\begin{tabular}{@{}cc@{}}
\toprule
Application Domain & Occurrence of Spatial Variability\\ 
\midrule
 Temperature & Regional Temperature Change Prediction
 \\ 
Soil Carbon Emission & Local Farmland Emission vs Global
\\
Precipitation & Regional variability in precipitation
\\
Extreme wind & Regional wind variability (e.g., wind farms)
\\
Climate Policy Making & County vs State level jurisdictions.
 \\
 \bottomrule
\vspace*{-5\abovedisplayskip}
\end{tabular}
\end{table}

The problem of statistical downscaling in climate data is challenging due to climate data's inherent complexity and variability involving intricate spatial and temporal dependencies influenced by numerous physical processes (e.g., ocean currents, wind patterns, temperature gradients, and coastal topography). The non-stationarity of climate data, driven by dynamic changes in climate patterns and extreme events, further complicates the development of reliable and robust downscaling techniques. For example, statistical properties like mean and standard deviation may not be constant over time \cite{liu2022downscaling, church2013sea}.

% To address these challenges, we used Kriging with Conditional Diffusion to effectively capture spatial variability and non-stationarity while preserving fine-scale features.

% Additionally, ensuring that downscaled models accurately reflect these dependencies while preserving physical constraints is difficult. The challenges in understanding the problem of statistical downscaling in sea-level rise data are threefold. Accurately capturing spatial dependencies is challenging due to the need to represent underlying physical processes such as ocean currents, wind patterns, temperature gradients, and coastal topography, all of which influence sea-level rise. Additionally, stationarity models, which assume that statistical properties like mean and standard deviation are constant over time, are unsuitable for sea-level rise data due to their dynamic nature. Furthermore, existing models often result in over-smoothing, which eliminates critical details and fails to preserve fine-scale features. 

Traditional Kriging and interpolation-based methods \cite{vicente2014improved, teutschbein2012evaluation, maraun2018statistical} for downscaling sea-level rise data often fall short due to their reliance on stationarity assumptions and their inability to capture the dynamic nature of climate data. These methods typically smooth the data excessively, resulting in the loss of critical fine-scale features. The loss leads to an inaccurate representation of the underlying physical processes, such as ocean currents, wind patterns, and temperature gradients. On the other hand, machine learning-based methods \cite{watt2024generative, leinonen2020stochastic, wan2024debias}, while powerful in capturing non-linear relationships, often struggle with generalization and maintaining physical consistency. These models can be data-hungry and sensitive to the quality and quantity of training data.

Traditional super-resolution models, designed primarily for image processing, cannot be directly applied to climate downscaling for the following reasons. (a) They lack physical constraints and fail to incorporate the complex, nonlinear relationships between various components of the Earth system \cite{oppenheimer2019sea, garner2018evolution}. (b) They do not handle the high-dimensional nature of climate data effectively and often require large amounts of high-resolution training data, typically sparse in climate science \cite{wang2021coastal}. (c) They are deterministic and do not provide probabilistic outputs or uncertainty estimates, which are crucial for climate predictions \cite{mcgranahan2007rising}. (d) They lack generalization across different regions and periods, as they do not account for the multiple scales inherent in climate processes \cite{kopp2014probabilistic}.

% Traditional super-resolution models, designed primarily for image processing, face additional challenges such as lack of physical constraints, inability to handle high-dimensional climate data, and failure to provide probabilistic outputs. 

To address these limitations, we propose a Kriging-informed Conditional Diffusion Probabilistic Model (Ki-CDPM), which combines the strengths of Kriging's spatial interpolation capabilities with the flexibility and robustness of conditional diffusion processes. Ki-CDPM effectively captures spatial variability, maintains physical consistency, handles high-dimensional data, and provides probabilistic outputs, ensuring more accurate and physically consistent downscaling of climate data.

% Adjusting the spacing before and after the table
\setlength{\intextsep}{2pt} % Adjust space above and below the table

\begin{table}[h]
    \centering
    \small % Reduce the font size
    \caption{Comparison of methods for sea-level downscaling}
    \begin{tabular}{|p{4.0cm}|c|c|}
        \hline
        \multicolumn{1}{|c|}{} & \multicolumn{2}{c|}{\textbf{ML based methods}} \\ \cline{2-3}
        \multicolumn{1}{|c|}{\textbf{}} & \textbf{No} & \textbf{Yes} \\ \hline
        \textbf{Kriging (spatial variability)} & & \\ \hline
        No & x & \cite{watt2024generative,leinonen2020stochastic,wan2024debias} \\ \hline
        Yes & \cite{vicente2014improved, teutschbein2012evaluation, maraun2018statistical} & \textbf{Ki-CDPM} \\ \hline
    \end{tabular}
    \label{tab:comparison}
\end{table}

\textbf{Contributions:} The paper contributions are as follows:
\begin{itemize}[noitemsep,topsep=0pt]
% \item We defined the problem of estimating fine-scale spatial patterns variability in regional sea-level rise downscaling.
\item We propose a novel Kriging-informed Conditional Diffusion Probabilistic Model (Ki-CDPM) that leverages Kriging interpolation for fine-scale projections.
% \item We introduce a modified U-Net architecture incorporating Kriging interpolated elevation maps at different scales to improve performance.
\item We introduce Variogram-Based Regularization to capture spatial variability in regional processes and enhance the physical consistency of downscaled data.
\item We compare the Ki-CDPM against state-of-the-art machine learning models and conduct comprehensive evaluations in different regions, demonstrating its effectiveness.
\end{itemize}
\textbf{Relevance to SIGSPATIAL:} This paper is relevant to SIGSPATIAL for the following reasons:
\begin{itemize}[noitemsep,topsep=0pt]
    \item The paper proposes a novel approach to addressing the challenge of downscaling climate data, which aligns with the fields of remote sensing, earth observation, spatial data mining, and knowledge discovery
    \item Models used in the paper (e.g., diffusion models) have been explored recently in SIGSPATIAL with promising results in graph forecasting~\cite{wen2023diffstg} and urban flow~\cite{zhou2023towards}.
    \item Challenges described in the paper (e.g., spatial variability) have been studied recently in SIGSPATIAL (e.g., ~\cite{lin2023modeling}).
    \item The paper integrates geostatistical methods and deep learning models, aligning with SIGSPATIAL's focus on GeoAI.
\end{itemize}

% This paper focuses on downscaling sea-level rise predictions, though the proposed Kriging-informed Conditional Diffusion Probabilistic Model (Ki-CDPM) is adaptable to other climate-related applications.
\textbf{Scope:} This paper focuses on the statistical downscaling of climate variables like sea-level rise and ocean eddy energy. Temporal downscaling is not addressed in this manuscript but will be explored in future work. This study does not explore physics-informed machine-learning methods for downscaling or does not simulate the temporal evolution of physical processes. Additional physical constraints can be added to the proposed method to explore downscaling climate variables with conservation properties.  

\textbf{Organization:} The paper is organized as follows: Section \ref{sec:Problem Definition} introduces basic concepts and provides the problem formulation. Section \ref{sec:Proposed Approach} describes the overall architecture of the Kriging-informed Conditional Diffusion Probabilistic Model (Ki-CDPM) and the variogram-based regularization. Experimental evaluations are presented for the proposed approach in Section \ref{sec:ExperimentalEvaluation}. Related work is mentioned in Section \ref{sec:Related_work}. Finally, Section \ref{section:conclusion_future_work} concludes this work and briefly lists the future work.

\section{Problem Formulation}
\label{sec:Problem Definition}
% and provides some domain-related background.
% In this subsection, we first define basic concepts and necessary tools for the problem statement.

\subsection{Basic Concepts}

\begin{definition}
A \textbf{Sea-Level Rise} or \textbf{Sea-Level Elevation Map} is a raster map that provides information on the average increase in the water level of the Earth's oceans. Sea level is the height of the sea surface relative to a standardized baseline, typically the mean sea level. It is measured using satellite altimetry, tide gauges, and GPS data, which require advanced algorithms for integrating and correcting various environmental factors. Understanding sea level elevation is crucial for climate research, disaster management, and urban planning and involves complex data processing, modeling, and predictive analytics \citep{hamlington2020understanding}.  
\end{definition}

Figure \ref{fig:Problem_Statement} shows sea-level rise information for the eastern equatorial Pacific Ocean and coastal Peru and Ecuador regions, where red and green contours depict high and low sea-level rise, respectively. From the coarse scale resolution (background color), we can observe that sea-level elevation along this coastal region was higher in August 2023 than in October 2023.

\begin{figure}[h]
    \centering
    \includegraphics[width=\columnwidth]{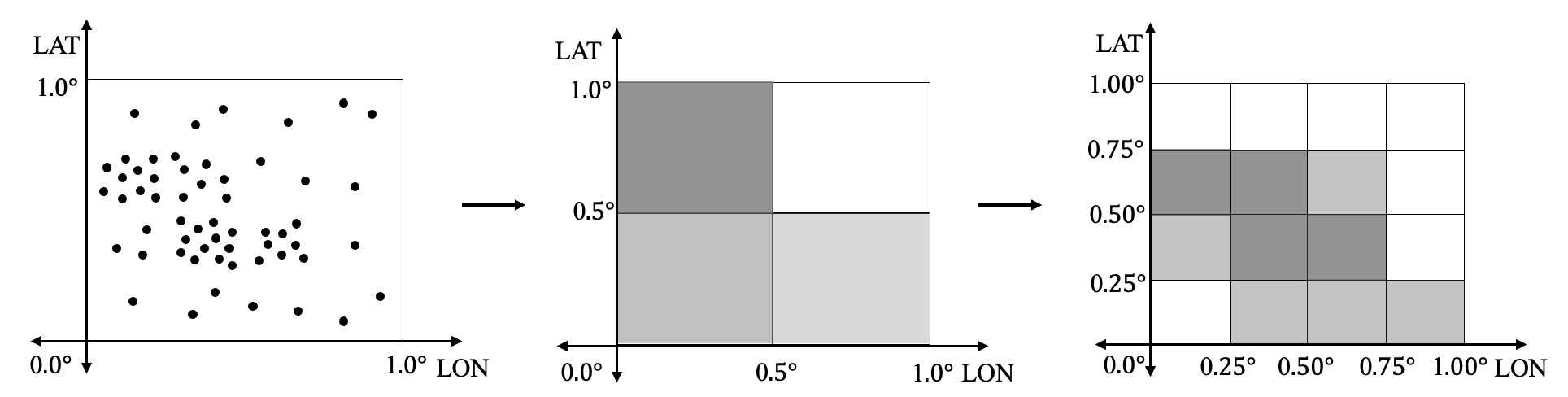}
    \caption{Coarse-scale and fine-scale resolution example.}
    \Description{An illustration of a coarse-scale and fine-scale resolution example.}
    \label{fig:Basic_Concept}
\end{figure}

\begin{definition}
A \textbf{coarse-scale} resolution is a $N \times N$ grid-based representation of a climate variable (e.g., sea-level elevation) for a given geographic region ($R$) projection.
\end{definition}

Figure \ref{fig:Basic_Concept} shows a study area with a $1^{\circ} \times 1^{\circ}$ spatial resolution, containing randomly distributed data points that were later discretized to a $0.5^{\circ} \times 0.5^{\circ}$ resolution, revealing variations in population density. 

\begin{definition}
A \textbf{fine-scale} resolution is an $M \times M$ (where M > N) grid-based representation of a climate variable (e.g., sea-level elevation) for a given geographic region ($R$) projection.
\end{definition}

Figure \ref{fig:Basic_Concept}(c) shows a study area with a spatial resolution of $0.25^{\circ} \times 0.25^{\circ}$ revealing finer scale variations in population density.

\begin{definition}
\textbf{Downscaling} in climate science derives fine-scale resolution data from coarse-resolution variables \citep{liu2022downscaling}. 
\end{definition}
% ,liu2016dynamical,hermans2020improving,kim2021local

\begin{figure}[h]
    \centering
    \includegraphics[width=\columnwidth]{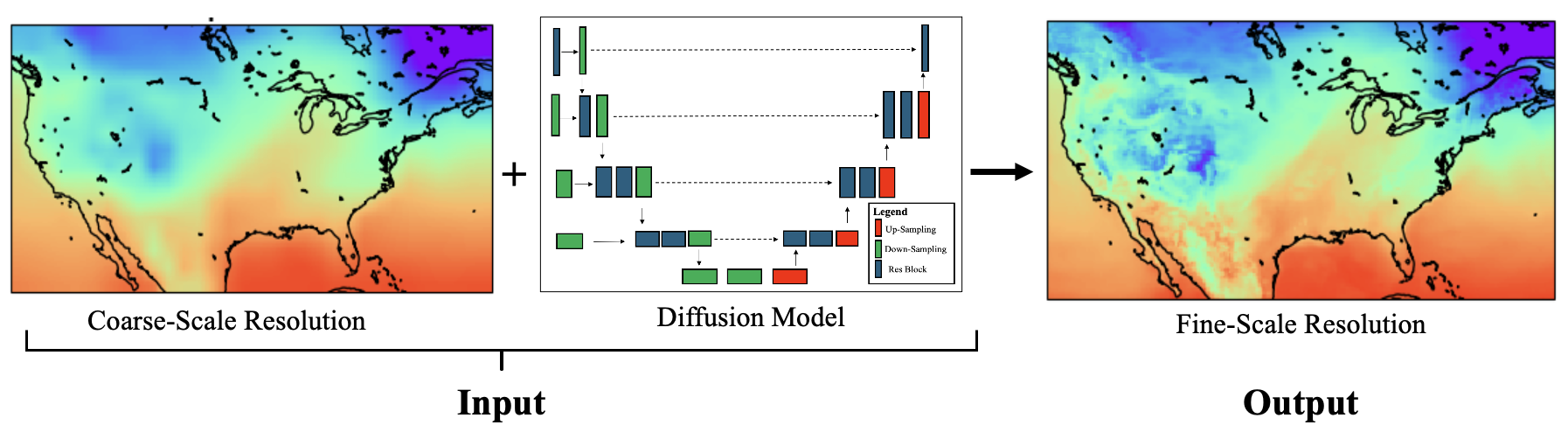}
    \caption{An illustrative example of input and output (Maps adapted from \cite{watt2024generative}).}
    \Description{An illustration showing the input and output maps adapted from Watt (2024).}
    \label{fig:Input_Output}
\end{figure}

Figure \ref{fig:Input_Output} shows the down-scaling process for a real-world temperature heatmap. Spatial variability within the map can be observed between the northern and the southern regions. Further, it is observed that the finer resolution data is better for observing regional patterns (e.g., Regions to the West of the center).

\subsection{Problem Formulation}\label{problem_definition}
The problem of downscaling a climate variable (e.g., sea-level elevation) is formally defined as follows:\\
\textbf{Input:}
\begin{enumerate}
	\item Coarse-Resolution climate data ($\Tilde{y}$)
        \item A diffusion model to output a downscaled climate dataset ($\Tilde{x}_{\text{high}}$) which is an approximation of the ground truth ($x_{\text{high}}$)

\end{enumerate}
\textbf{Output:} High-resolution downscaled climate data ($\Tilde{x}_{\text{high}}$) as an approximation of the ground truth data ($x_{\text{high}}$) \\
\textbf{Objective:} Solution quality, model generalization\\
\textbf{Constraints:}
(1) Spatial variability; (2) Domain adaptability; (3) Model interpretability\\

% \begin{figure}[htbp]
%     \centering
%     \begin{subfigure}{0.65\columnwidth}
%         \includegraphics[width=\linewidth]{Images/Toy Input.png}
%         \caption{$0.5^{\circ} \times 0.5^{\circ}$ Resolution}
%         \label{fig:Input1}
%     \end{subfigure}%
%     \begin{subfigure}{0.40\columnwidth}
%         \includegraphics[width=\linewidth]{Images/Real World Input.png}
%         \caption{Coarse-scale Resolution}
%         \label{fig:Input2}
%     \end{subfigure}
%     \caption{Problem Statement Input}
%     \label{fig:Input}
% \end{figure}
% \vspace{-1.5em}
% \begin{figure}[htbp]
%     \begin{subfigure}{0.65\columnwidth}
%         \includegraphics[width=\linewidth]{Images/Toy Output.png}
%         \caption{$0.25^{\circ} \times 0.25^{\circ}$ Resolution}
%         \label{fig:Output1}
%     \end{subfigure}%
%     \begin{subfigure}{0.40\columnwidth}
%         \includegraphics[width=\linewidth]{Images/Real World Output.png}
%         \caption{Fine-scale Resolution}
%         \label{fig:Output2}
%     \end{subfigure}
%     \caption{Problem Statement Output}
%     \label{fig:Output}
% \end{figure}

\textbf{Inverse Problem:} Downscaling sea-level rise projections from coarse-resolution climate models to finer regional scales can be thought of as an inverse problem aimed at estimating high-resolution sea-level changes based on limited low-resolution observations and a forward model \cite{liu2022downscaling, jiang2020downscaling}. This problem is complex due to nonlinear relationships between inputs and outputs, sparse and unevenly distributed observations, and model errors and uncertainties \cite{church2013sea, slangen2014projecting, garner2018evolution, oppenheimer2019sea}. Additionally, the high-dimensional nature of the problem poses computational challenges \cite{kopp2014probabilistic}. These factors make sea-level rise data downscaling an ill-posed inverse problem, which requires advanced techniques for reliable high-resolution projections \cite{wang2021coastal, mcgranahan2007rising}.

\section{Kriging-informed Conditional Diffusion Probabilistic Model}
\label{sec:Proposed Approach}
This section will introduce a novel architecture based on Kriging interpolation as a conditional input to preserve spatial variability while transforming coarser-resolution climate variables (e.g., sea-level elevation) to finer-scale resolution. Section \ref{sec:details}
provides a detailed explanation of the proposed architecture, and Section \ref{sec:regularizer} shows the Kriging-informed training and regularization.

\subsection{Conditional Diffusion Probabilistic Model}
\label{subsection:ConditonalDiffusion}
% $\mathcal{D} = \{\vx_i\}_{i=1}^M$
In this approach, the goal is to generate a fine-scale resolution map from a coarse-resolution input map in which samples were drawn from an unknown conditional distribution $p(\vx \,|\, \vy)$ where $p(\vy)$ is a distribution of Kriging-interpolated map of the coarse-scale resolution input ($\Tilde{y}$). The objective is to learn a parametric approximation of $p(\vx \,|\, \vy)$ via {\em stochastic} refinements which iteratively maps source condition $\vy$ to a target output $\vx \in \R^d$ via denoising diffusion probabilistic (DDPM) model ~\cite{ho2020denoising,sohl2015deep}.

\begin{figure}[htb]
    \centering
    \includegraphics[width=0.475\textwidth]{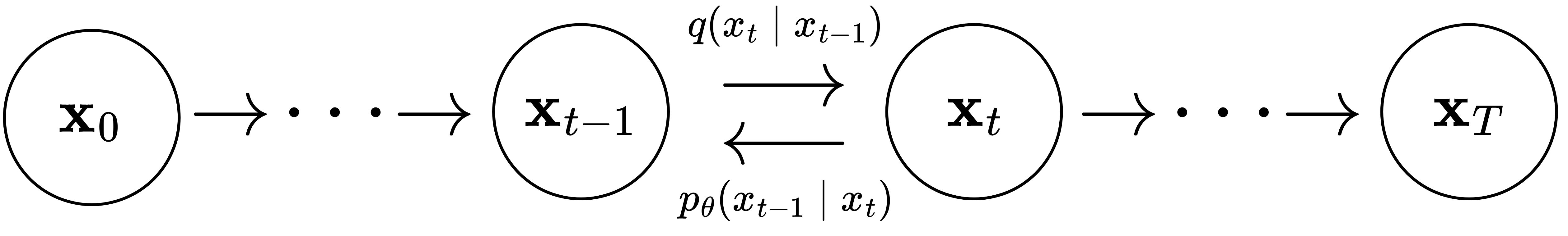}
    \captionsetup{justification=centering}
    \caption{Forward and Reverse Diffusion Processes}
    \Description{A diagram illustrating the forward and reverse diffusion processes.}
    \label{fig:DiffusionToy}
\end{figure}

Figure \ref{fig:DiffusionToy} presents a detailed view of a diffusion model where isotropic Gaussian noise is progressively added to a signal through a predetermined Markov chain, denoted by $q(\vx_t \mid \vx_{t-1})$. This process generates a series of intermediate noisy maps referred to as \textit{forward} diffusion. Conversely, in \textit{reverse} diffusion, the process begins with a completely noisy map, $\vx_T \sim \mathcal{N}(\bm{0}, \bm{I})$. The model then incrementally refines this map through successive iterations $(\vx_{T-1}, \vx_{T-2}, \dots, \vx_0)$ using \textbf{learned conditional transition distributions} $p_\theta(\vx_{t-1} \mid \vx_t, \vy)$, aiming for $\vx_0 \sim p(\vx \mid \vy)$. This method reverses the forward diffusion process by reconstructing the signal from noise using a backward Markov chain conditioned on $\vy$.
% Further details on forward and reverse diffusion, cited in \cite{ho2020denoising,sohl2015deep}, are provided in Appendix A.

% While in principle, this diffusion process can also condition on the source image $\vy$, \eg~$q(\vx_t \,|\, \vx_{t-1}, \vx)$, in super-resolution, all of the information of $\vy$ is already contained in $\vx_0$, making the diffusion process unconditional on $\vy$ is a reasonable choice. 

% We learn the reverse chain using a neural denoising model $f_\theta$
% that takes a source image and a noisy target image as input and estimates the noise.
%In what follows, 
% We first give an overview of the forward diffusion process,
% and then discuss how our denoising model $f_\theta$ is trained and used for inference.

\textbf{Forward Process ($q$):} In this process, gaussian noise is progressively added to a fine-scale resolution $\vx_0$ over $T$ iterations \cite{ho2020denoising,sohl2015deep}: 
% In a \textit{forward} Markovian diffusion \cite{ho2020denoising,sohl2015deep} process $q$, which incrementally introduces Gaussian noise into a fine-scale resolution $\vx_0$ across $T$ iterations:
\begin{eqnarray}
  q(\vx_{1:T} \mid \vx_0) &=& \prod\nolimits_{t=1}^{T} q(\vx_{t} \mid \vx_{t-1})~, \\
  q(\vx_{t} \mid \vx_{t-1}) &=& \mathcal{N}(\vx_{t} \mid \sqrt{\alpha_t}\, \vx_{t-1}, (1 - \alpha_t) \bm{I} )~, %\vy_{t}; 
\end{eqnarray}
Here, the hyperparameters $\alpha_{1:T}$ are constrained between $0$ and $1$, representing the noise variance introduced at each iteration. The variable $\vx_{t-1}$ is scaled down by a factor of $\sqrt{\alpha_t}$ to keep the variance of the random variables finite. Additionally, deriving $\vx_t$ from $\vx_0$ can be streamlined using Equation \ref{eq:diffusion-marginalized}.:
% where the scalar parameters $\alpha_{1:T}$ are hyper-parameters, subject to $0 < \alpha_t < 1$, determining the noise variance added at each iteration. Here, $\vx_{t-1}$ is scaled down by $\sqrt{\alpha_t}$ to keep the variance of the random variables finite as $t$ approaches infinity. Therefore, the intermediate steps for determining $\vx_t$ from $\vx_0$ can be simplified using Equation \ref{eq:diffusion-marginalized}:
% Note that $\vx_{t-1}$ is attenuated by $\sqrt{\alpha_t}$ to ensure that the variance of the random variables remains bounded as $t \to \infty$. Hence, the intermediate steps to categorize $\vx_t$ given $\vx_0$ can be marginalized by Equation \ref{eq:diffusion-marginalized}:
% For instance, if the variance of $\vx_{t-1}$ is $1$, then the variance of $\vx_{t}$ is also $1$. 
\begin{equation}
    q(\vx_t \mid \vx_0) ~=~ \mathcal{N}(\vx_t \mid \sqrt{\psi_t}\, \vx_0, (1-\psi_t) \bm{I})~,
\label{eq:diffusion-marginalized}
\end{equation}
where $\psi_t = \prod_{i=1}^t \alpha_i$. In addition,  
one can derive the posterior distribution of $\vx_{t-1}$ given $(\vx_0, \vx_t)$ as
\begin{equation}
\begin{aligned}
    q&(\vx_{t-1} \mid \vx_0, \vx_t) = \mathcal{N}(\vx_{t-1} \mid \bm{\mu}, \sigma^2 \bm{I})\\
    &\bm{\mu} =  \frac{\sqrt{\psi_{t-1}}\,(1-\alpha_t)}{1-\psi_t}\, \vx_0 \!+\! \frac{\sqrt{\alpha_t}\,(1-\psi_{t-1})}{1-\psi_t}\vx_t\\
    &\sigma^2 = \frac{(1-\psi_{t-1})(1-\alpha_t)}{1-\psi_t}~.
\end{aligned}
\label{eq:posteriror-ytm1}
\end{equation}
% This posterior distribution is helpful when parameterizing the reverse chain and formulating a variational lower bound on the log-likelihood of the reverse chain.
% We next discuss learning a neural network to reverse this Gaussian diffusion process.

% \algrenewcommand\algorithmicindent{0.5em}%
% \begin{figure}[t]
% \vspace*{-.5cm}
% \small
% \begin{minipage}[t]{0.495\textwidth}
% \begin{algorithm}[H]
%   \caption{Training} \label{alg:training}
%   \small
%   \begin{algorithmic}[1]
%     \Repeat
%       \State $\bx_0 \sim q(\bx_0)$
%       \State $t \sim \mathrm{Uniform}(\{1, \dotsc, T\})$
%       \State $\bepsilon\sim\mathcal{N}(\bzero,\bI)$
%       \State Take gradient descent step on
%       \Statex $\qquad \grad_\theta \left\| \bepsilon - f_\theta(\sqrt{\bar\alpha_t} \bx_0 + \sqrt{1-\bar\alpha_t}\bepsilon, y) \right\|^p_$
%     \Until{converged}
%   \end{algorithmic}
% \end{algorithm}
% \end{minipage}
% \vspace*{-.2cm}
% \end{figure}

\algrenewcommand\algorithmicindent{0.5em}%
\begin{figure}[t]
\vspace*{-.5cm}
\small
\begin{minipage}[t]{0.495\textwidth}
\begin{algorithm}[H]
  \caption{Training} \label{alg:training}
  \small
  \begin{algorithmic}[1]
    \Repeat
      \State $(\vx_0, \vy) \sim p(\vx, \vy)$
      \State $\psi \sim p(\psi)$
      \State $\bm{\epsilon}\sim\mathcal{N}(\mathbf{0},\mathbf{I})$
      \State Take a gradient descent step on
      \Statex $\qquad \nabla_\theta \left\lVert \veps - f_\theta(\vy, \sqrt{\psi} \vx_0 + \sqrt{1-\psi} \bm{\epsilon}, \psi) \right\rVert_p^p$ 
    \Until{converged}
  \end{algorithmic}
\end{algorithm}
\end{minipage}
\Description{An algorithm that describes the training process for a model, involving sampling from distributions, applying gradient descent, and optimizing the model until convergence.}
\vspace*{-.2cm}
\end{figure}

\textbf{Reverse Process ($p_\theta$):} 
\label{sec:inference}
In this process, we leverage additional kriging interpolated source elevation map $\vy$ and optimize a neural denoising model $f_{\theta}$ that takes as input map y and a noisy map $\Tilde{x}$,
\begin{equation}
    \Tilde{x} = \sqrt{\psi}\, \bm{x}_0 + \sqrt{1-\psi} \,\xeps~,
    ~~~~~~ \veps \sim \mathcal{N}(\bm{0},\bm{I})~,
\label{eq:noisy-y}
\end{equation}
% where $\veps \sim \mathcal{N}(\bm{0}, \bm{I})$,
Equation \ref{eq:noisy-y} aims to restore the noise-free target map $\bm{x}_{0}$, and $\Tilde{x}$ aligns with the marginal distribution of noisy maps at various stages of the forward diffusion process described in \eqref{eq:diffusion-marginalized}.
% Instead of regressing the output of $f_{\theta}$ to $\veps$, as in \eqref{eq:loss}, one can also regress the output of $f_{\theta}$ to $\vy_0$.
% Given $\gamma$ and $\vty$, the values of $\veps$ and $\vy_0$ can be derived from each other deterministically,
% but changing the regression target impacts the scale of the loss function.
% We expect both of these variants to work reasonably well if $p(\gamma)$ is modified to account for the scale of the loss function.
% Further investigation of the loss function used for training the denoising model is an interesting avenue for future research.

% \subsubsection{Inference via Iterative Refinement}
% \label{sec:inference}

In contrast to the forward process $q$, $p_\theta$ goes in the reverse direction starting from Gaussian noise $\bm{x}_{T}$:
\begin{eqnarray}
    p_\theta(\vx_{0:T} | \vy) &=& p(\vx_T) \prod\nolimits_{t=1}^T p_\theta(\vx_{t-1} | \vx_t, \vy) \\
    p(\vx_T) &=& \mathcal{N}(\vx_T \mid \bm{0}, \bm{I}) \label{eq:mutheta}\\
    p_\theta(\vx_{t-1} | \vx_{t}, \vy) &=& \mathcal{N}(\vx_{t-1} \mid \mu_{\theta}(\vy, {\vx}_{t}, \psi_t), \sigma_t^2\bm{I})~. \label{eq:reverse_process}
\end{eqnarray}
The inference process is defined using isotropic Gaussian conditional distributions, $p_\theta(\vx_{t-1} | \vx_{t}, \vy)$, that are learned. If the noise variance in the forward process steps is minimized, \emph{i.e.}, $\alpha_{1:T} \approx 1$, the resulting optimal reverse process $p(\vx_{t-1} | \vx_{t}, y)$ will closely approximate a Gaussian distribution~\cite{sohl2015deep}. The Gaussian conditionals selected for the inference process in \eqref{eq:reverse_process} can closely approximate the actual reverse process. Moreover, it is necessary for $1 - \psi_T$ to be large enough to ensure that the distribution of $\mathbf{x}_T$ closely aligns with the prior distribution $p(\mathbf{x}_T) = \mathcal{N}(\mathbf{x}_T | \mathbf{0}, \mathbf{I})$, which is a standard Gaussian distribution with mean zero and identity covariance matrix. Here we designed $f_\theta$ to predict $\veps$ from any noisy map $\Tilde{x}$, including $x_t$. Consequently, we can estimate $\vx_0$ by rearranging the terms as shown in \eqref{eq:noisy-y}:
% We define the inference process in terms of isotropic Gaussian conditional distributions, $p_\theta(\vx_{t-1} | \vx_{t}, \vy)$, which are learned.
% If the noise variance of the forward process steps is set as small as possible, \ie~$\alpha_{1:T} \approx 1$, the optimal reverse process $p(\vx_{t-1} | \vx_{t}, y)$ will be approximately Gaussian~\cite{sohl2015deep}.
% The chosen Gaussian conditionals in the inference process  \eqref{eq:reverse_process} can adequately approximate the actual reverse process. Additionally, the value of $1 - \psi_T$ needs to be sufficiently large to ensure that the distribution of $\mathbf{x}_T$ closely resembles the prior distribution  $p(\mathbf{x}_T) = \mathcal{N}(\mathbf{x}_T | \mathbf{0}, \mathbf{I})$, which is a standard Gaussian distribution with mean zero and identity covariance matrix. 
% The denoising model $f_\theta$ is trained to estimate $\veps$, given any noisy map $\Tilde{x}$ including $x_{t}$. Thus, given $\vx_t$, we approximate $\vx_0$ by rearranging the terms in \eqref{eq:noisy-y} as
\begin{equation}
    \hat{\vx}_0 = \frac{1}{\sqrt{\psi_t}} \left( \vx_t - \sqrt{1 - \psi_t}\, f_{\theta}(\vy, \vx_{t}, \psi_t) \right)~.
\end{equation}

% Estimated $\hat{\vx}_0$ is further substituted into the posterior distribution of $q (\vx_{t-1} | \vx_0, \vx_t)$ to parameterize mean of $p_\theta(\vx_{t-1} | \vx_t, \vy)$ as
Finally, we insert $\hat{\vx}_0$ into the posterior distribution of $q (\vx_{t-1} | \vx_0, \vx_t)$, which parameterizes the mean of $p_\theta(\vx_{t-1} | \vx_t, \vy)$ in Equation \ref{eq:12}. We set the variance of $p_\theta(\vx_{t-1}|\vx_t, \vy)$ to $(1 - \alpha_t)$, which is the default variance as determined by the variance of the forward process \cite{ho2020denoising}:
\begin{equation}
\label{eq:12}
    \mu_{\theta}(\vy, {\vx}_{t}, \psi_t) = \frac{1}{\sqrt{\alpha_t}} \left( \vx_t - \frac{1-\alpha_t}{ \sqrt{1 - \psi_t}} f_{\theta}(\vy, \vx_{t}, \psi_t) \right)~,
\end{equation}
% and we set the variance of $p_\theta(\vx_{t-1}|\vx_t, \vy)$ to $(1 - \alpha_t)$, a default given by the variance of the forward process \cite{ho2020denoising}.

% Following this parameterization, each iteration of iterative refinement under our model takes the form,
Additionally, we carry out iterative refinement at each iteration as described in Equation \ref{eq:13} (where $\veps_t \sim \mathcal{N}(0,1)$):
\begin{equation}
\label{eq:13}
\vx_{t-1} \leftarrow \frac{1}{\sqrt{\alpha_t}} \left( \vx_t - \frac{1-\alpha_t}{ \sqrt{1 - \psi_t}} f_{\theta}(\vy, \vx_{t}, \psi_t) \right) + \sqrt{1 - \alpha_t}\veps_t~,
\end{equation}
% where $\veps_t \sim \stdnormal$. This resembles one step of Langevin dynamics, which provides an estimate of the gradient of the data log density.

\begin{figure}
\vspace*{-.5cm}
\small
\begin{minipage}[t]{0.495\textwidth}
\begin{algorithm}[H]
  \caption{Inference} \label{alg:sampling}
  \small
  \begin{algorithmic}[1]
    \vspace{.04in}
    \State $\vx_T \sim \mathcal{N}(\mathbf{0}, \mathbf{I})$
    \For{$t=T, \dotsc, 1$}
      \State $\vz \sim \mathcal{N}(\mathbf{0}, \mathbf{I})$ if $t > 1$, else $\vz = \mathbf{0}$
      \State $\vx_{t-1} = \frac{1}{\sqrt{\alpha_t}}\left( \vx_t - \frac{1-\alpha_t}{\sqrt{1-\psi_t}} f_\theta(\vx_t, \vy, \psi_t) \right) + \sqrt{1 - \alpha_t} \vz$
    \EndFor
    \State \textbf{return} $\vx_0$
    \vspace{.04in}
  \end{algorithmic}
\end{algorithm}
\end{minipage}
\Description{An algorithm for the inference process, starting from a random sample from a normal distribution and iteratively updating the sample using model parameters until reaching the final output.}
\vspace*{-0.2cm}
\end{figure}

\subsection{Interpolation with Kriging}
Universal Kriging (U-Krig) is an advanced geostatistical method that allows for interpolation by accounting for deterministic trends and spatial correlation in the data. For instance, let \(\Tilde{y}\) represent a coarser resolution map and \(y\) represent a finer resolution map. We then model the trend using a second-order polynomial function of the spatial coordinates:

\begin{align}
	m(s) = \beta_0 + \beta_1 x + \beta_2 y + \beta_3 x^2 + \beta_4 y^2 + \beta_5 xy
\end{align}

Where \(x\) and \(y\) are the spatial coordinates (longitude and latitude), and \(\beta_0, \beta_1, \beta_2, \beta_3, \beta_4, \beta_5\) are the coefficients to be estimated. Hence, the residuals \(\gamma(s_i)\) are calculated by subtracting the estimated trend \(m(s)\) from the observed sea-level elevations \(\Tilde{y}(s_i)\):

\begin{align}
	\gamma(s_i) = \Tilde{y}(s_i) - m(s_i)
\end{align}

These residuals represent the deviations from the deterministic trend and are used to model the spatial correlation structure. 

We then use the spatial correlation of the residuals, which is quantified using the \textit{experimental} variogram. The semivariance \(\Delta(h)\) for different lag distances \(h\) is computed as:

\begin{align}
	\Delta(h) = \frac{1}{2N(h)} \sum_{i=1}^{N(h)} [\gamma(s_i) - \gamma(s_i + h)]^2
\end{align}

Where \(N(h)\) is the number of pairs of points separated by a distance \(h\), and \(\gamma(s_i)\) are the residuals. The experimental variogram is fitted with a theoretical Matérn variogram model, defined as:

\begin{align}
	\Delta(h) = \sigma^2 \left[ 1 - \frac{2^{1-\nu}}{\Gamma(\nu)} \left( \frac{h}{\rho} \right)^\nu K_\nu \left( \frac{h}{\rho} \right) \right] + C_0
\end{align}

where \(\nu\) is the smoothness parameter, \(\rho\) is the range parameter, \(\sigma^2\) is the partial sill, \(C_0\) is the nugget, and \(K_\nu\) is the modified Bessel function of the second kind.
Once the variogram parameters ($\nu, \rho, \sigma^2, C_0$) are estimated by fitting the Matérn variogram model to the experimental variogram, they are used to calculate the semivariances $\Delta(s_i - s_j)$ between all pairs of sampled locations. These semivariances form the basis of the Kriging system of equations.
We use the Universal Kriging system to predict the sea-level elevation at unsampled locations with a higher resolution. The predicted value \(y(s_0)\) at location \(s_0\) is given by:

\begin{align}
	y(s_0) = \sum_{i=1}^{n} \lambda_i \Tilde{y}(s_i) + \sum_{k=1}^{p} \mu_k f_k(s_0)
\end{align}

where \(\lambda_i\) are the Kriging weights, \(\mu_k\) are the Lagrange multipliers for the trend functions, and \(f_k(s_0)\) are the basis functions for the trend model. The weights \(\lambda_i\) are obtained by solving the system of equations:
\[
\begin{cases}
\sum_{j=1}^{n} \lambda_j \Delta(s_i - s_j) + \sum_{k=1}^{p} \mu_k f_k(s_i) = \Delta(s_i - s_0) & \text{for } i = 1, \ldots, n \\
\sum_{j=1}^{n} \lambda_j f_k(s_j) = f_k(s_0) & \text{for } k = 1, \ldots, p
\end{cases}
\]
where \(\Delta(s_i - s_j)\) is the semivariance between locations \(s_i\) and \(s_j\). Finally, we apply the Universal Kriging weights to interpolate the sea-level elevation data to a higher resolution. This interpolated map \(y\) is used as a conditional input to the diffusion model, which then controls the generation in the reverse diffusion process to retain the spatial dependencies.

\subsection{\modelname Proposed Model Architecture}
\label{sec:details}

% Ki-CDPM leverages a stochastic process that generates multiple high-resolution sea-level elevation maps from a single coarse-resolution input by adding random noise (i.e., isotropic Gaussian) to the predicted denoised maps at each reverse diffusion step. For instance, let $\boldsymbol{\epsilon} \sim \mathcal{N}(\mathbf{0}, \mathbf{I})$ be a random noise tensor sampled from a standard Gaussian distribution with the \textbf{same spatial dimensions} as the high-resolution map. At each reverse diffusion step $t$, the predicted denoised map $\mathbf{x}_{t-1}$ is perturbed by adding the scaled noise tensor:
% \begin{equation}
% \mathbf{x}_{t-1} \leftarrow \mathbf{x}_{t-1} + \sigma_t \boldsymbol{\epsilon}
% \end{equation}
% where $\sigma_t$ is a scaling factor that controls the magnitude of the noise added at step $t$. The scaling factor can be chosen based on the desired level of stochasticity and the characteristics of the sea-level elevation data. Introducing random perturbations at each reverse diffusion step generates high-resolution maps that capture the uncertainties and variabilities in the downscaling process. Each generated map represents a plausible realization of the high-resolution sea-level elevations consistent with the coarse-resolution input and the estimated variogram parameters. Such maps can be used to assess the robustness of the downscaling results and quantify the uncertainties associated with the sea-level projections. 

The Ki-CDPM extends the CDPM by incorporating a conditioned input obtained from Universal-Kriging (U-Krig) on the coarse-resolution elevation map $\mathbf{\Tilde{y}} \in \mathbb{R}^{M \times M}$ providing local variability for the climate variable. The objective is to find an interpolated elevation map $\mathbf{y} \in \mathbb{R}^{N \times N}$ (where $N > M$) with the exact resolution as $x_{0}$. Hence, $\mathbf{y}$ is later used as a conditional input concatenated with the noisy elevation map $\mathbf{x}_t \in \mathbb{R}^{N \times N}$ at each diffusion step $t$ along the channel dimension, forming a multi-channel input tensor $[\mathbf{x}_t, \mathbf{y}] \in \mathbb{R}^{N \times N \times 2}$.

The concatenated input tensor is fed into the U-Net \cite{ronneberger2015u} architecture of the Ki-CDPM, allowing the model to learn the conditional distribution $p_{\theta}(\mathbf{x}_{t-1} | \mathbf{x}_t, \mathbf{y})$. The U-Net can leverage the spatially variable information provided by the U-Krig-based elevation map $\mathbf{y}$ to guide the generation of a fine-scale resolution map capturing spatial dependencies via leveraging the strengths of geostatistical information into the diffusion modeling framework. 

\begin{figure*}[ht]
  \centering
  \begin{subfigure}{0.49\textwidth}
    \includegraphics[width=\textwidth]{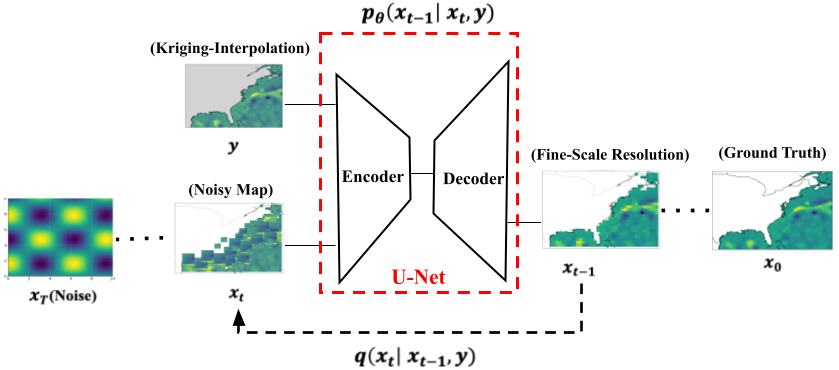}
    \caption{Conditional Diffusion Model Architecture}
    \label{fig:ConditionalDiffusion}
    \Description{A diagram illustrating the architecture of a Conditional Diffusion Model.}
  \end{subfigure}
  \hfill
  \begin{subfigure}{0.50\textwidth}
    \includegraphics[width=\textwidth]{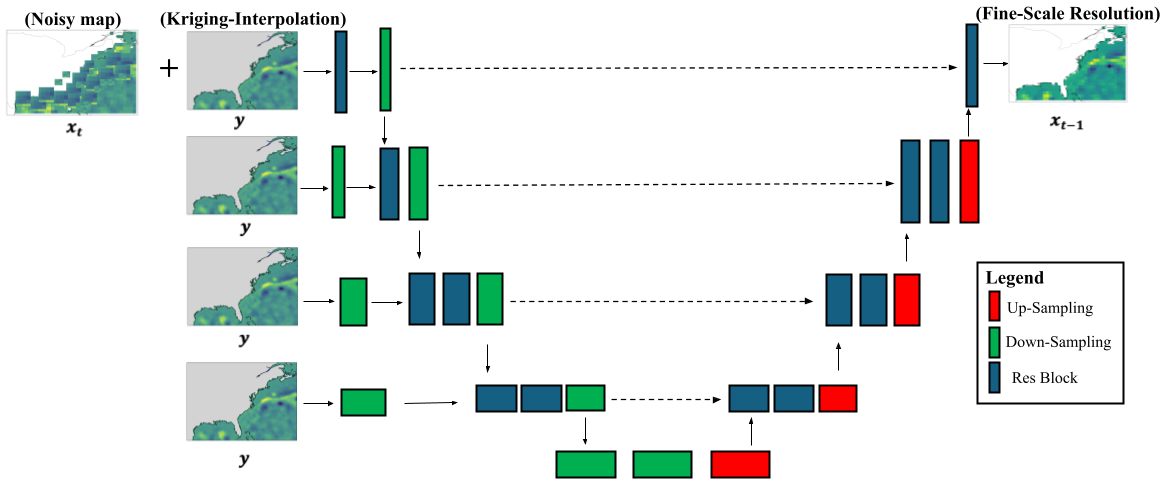}
    \caption{U-Net Model Architecture}
    \label{fig:UNet}
    \Description{A diagram showing the U-Net Model Architecture used in the Kriging-informed Conditional Diffusion Model.}
  \end{subfigure}
  \caption{Proposed Kriging-informed Conditional Diffusion Model (Ki-CDPM) Architecture}
  \Description{A figure showing two subfigures: one representing the Conditional Diffusion Model architecture and the other representing the U-Net architecture used in the model.}
  \label{fig:Architecture}
\end{figure*}

\textbf{Execution Trace:} Figure \ref{fig:Architecture} provides an overview of our proposed Ki-CDPM, a single conditional diffusion model that utilizes Kriging interpolated elevation map as conditional input $\mathbf{y}$. The forward process begins by adding small Gaussian noise until the input map is completely distorted. In \textit{reverse} process, $\bm{x}_{T}$ (i.e., complete random noise) is denoised until we recover $\Tilde{x}_{0}$( an approximation of $x_{0}$) and within each transition interval, we learn U-Net parameters which allow conditioning with Kriging-interpolated elevation map $y$ to generate finer-scale resolution map. For instance, Figure \ref{fig:Architecture} (a) shows a transition interval $x_{t}$ and $x_{t-1}$ where the \textit{forward} process is $\in$ $q(\mathbf{x}_{t} | \mathbf{x}_{t-1})$ and the \textit{reverse} process is $\in$ $p_{\theta}(\mathbf{x}_{t-1} | \mathbf{x}_t, \mathbf{y})$. At time step $t$, we concatenate the noisy map ($x_{t}$) with the U-Krig interpolated map ($y$) as a conditional input (where $y$ $\in$ $\mathbb{R}^{N \times N}$) and pass it to the U-Net (as shown in Figure \ref{fig:Architecture} (b)). Thus the \textit{reverse} process is represented by $p_{\theta}(\mathbf{x}_{t-1} | \mathbf{x}_t, \mathbf{y})$. In the U-net encoder, we utilize different scales of the conditional input ($y$) via several stacked down-sampling layers. The U-Net outputs $x_{t-1}$, a less noisy version of the elevation map. Finally, this process yields the denoised version or a more fine-scale map as the output. For our training noise schedule, we use a piecewise distribution~\cite{chen2021wavegrad}, i.e., $p(\psi) = \sum_{t=1}^T \frac{1}{T} U(\psi_{t-1}, \psi_t)$, where we first uniformly sample a time step $t \sim {0, \ldots, T}$, followed by $\psi \sim U(\psi_{t-1}, \psi_t)$ with $T = 1000$.

\subsection{Proposed Variogram-based Regularization}
\label{sec:regularizer}

\textbf{Variational Lower Bound ($\mathcal{L}_{\text{VLB}}$)}: Considering the forward diffusion process as a set approximate posterior within the inference mechanism, it is possible to establish the ensuing variational lower bound for the marginal log-likelihood:

\begin{align}
    \mathbb{E}_{(x_0,y)} \log p_\theta (x_0|y) &\geq \mathbb{E}_{y,x_0} \mathbb{E}_{q(x_{1:T}|x_0)} \bigg[ \log p(x_T) \nonumber \\
    &\quad + \sum_{t \geq 1} \log \frac{p_\theta(x_{t-1}|x_t, y)}{q(x_t|x_{t-1})} \bigg]
\end{align}

According to the specific parameterization of the inference process described earlier, the negative variational lower bound can be simplified and expressed as a loss function. This simplified loss consists of terms corresponding to each time step, and a constant factor weights each term.

\begin{align}
    \mathcal{L}_{\text{VLB}} = \mathbb{E}_{x, y_0, \epsilon} \left[ \sum_{t=1}^T \frac{1}{T} \left\| \epsilon - \epsilon_\theta \left(x, \sqrt{\psi_t y_0} + \sqrt{1 - \psi_t} \epsilon, \psi_t \right) \right\|_2^2 \right]
\end{align}

In this equation, $\epsilon$ represents a random variable that follows a standard normal distribution with mean 0 and identity covariance matrix $I$. It's important to note that this objective function is equivalent to the $L_2$ norm. Additionally, the distribution $p(\psi)$ is characterized as a uniform distribution over {$\psi_{1},...,\psi_{T}$}.\\

\textbf{Variogram-based Regularization:} To further exhibit Ki-CDPM towards spatial dependence structure as the observed finer-resolution data ($x_{0}$), we introduce \textbf{variogram-based regularization} in conjunction to Conditional Variational Lower Bound ($\mathcal{L}_{\text{VLB}}$). The regularization term $\mathcal{R}_V$ penalizes the discrepancy between the \textit{empirical} variogram of the generated noisy elevation maps (in the reverse process) and the variogram from the observed high-resolution map. 

Let $\Delta_{\bx_t}(\bm{h})$ denote the empirical variogram of the generated elevation map $\bm{x}_t$ at step $t$, calculated as:

\begin{equation}
\Delta_{\bx_t}(\bm{h}) = \frac{1}{2|\mathcal{N}(\bm{h})|} \sum_{(\bm{s}_i, \bm{s}_j) \in \bm{N}(\bm{h})} (x_t(\bm{s}_i) - \bm{x}_t(\bm{s}_j))^2
\end{equation}

where $\mathcal{N}(\bm{h})$ is the set of location pairs separated by the lag vector $\mathbf{h}$, and $|\cdot|$ denotes the cardinality of a set.

Let $\Delta_{\bx_0}(\bm{h})$ denote the variogram of the observed map ($x_{0}$) evaluated at lag $\mathbf{h}$ :

\begin{equation}
\Delta_{\bx_0}(\bm{h}) = \frac{1}{2|\mathcal{N}(\bm{h})|} \sum_{(\bm{s}_i, \bm{s}_j) \in \bm{N}(\bm{h})} (x_0(\bm{s}_i) - \bm{x}_0(\bm{s}_j))^2
\end{equation}

The variogram-based regularization term $\mathcal{R}_V$ is defined as the mean squared error between the empirical and observed variograms over a set of representative lag vectors $\mathcal{H}$:

\begin{equation}
\mathcal{R}_V = \frac{1}{|\mathcal{H}|} \sum_{\mathbf{h} \in \mathcal{H}} (\Delta_{\mathbf{x}_t}(\mathbf{h}) - \Delta_{\boldsymbol{\bm{x}_{0}}}(\mathbf{h}))^2
\end{equation}

The regularization term $\mathcal{R}_V$ is added to the original loss function of the CDPM, weighted by a hyperparameter $\lambda_V$:

\begin{equation}
\mathcal{L}_{\text{Ki-CDPM}} = \mathcal{L}_{\text{VLB}} + \lambda_V \mathcal{R}_V
\end{equation}

The hyperparameter $\lambda_V$ controls the strength of the variogram-based regularization and can be tuned to balance the trade-off between data fidelity and spatial structure preservation.

By minimizing the augmented loss function $L_{\text{Ki-CDPM}}$, the Ki-CDPM is encouraged to generate high-resolution elevation maps that match the finer-resolution observations and exhibit similar spatial dependence structures. Algorithm \ref{alg:training2} provides training for Ki-CPDM in detail.

\algrenewcommand\algorithmicindent{0.5em}%
\begin{figure}[t]
\vspace*{-.5cm}
\small
\begin{minipage}[t]{0.495\textwidth}
\begin{algorithm}[H]
  \caption{Training in Ki-CDPM} \label{alg:training2}
  \small
  \begin{algorithmic}[1]
    \Repeat
      \State $(\vx_0, \Tilde{\vy}) \sim p(\vx, \Tilde{\vy})$
      \State $t \sim \mathrm{Uniform}(\{1, \dotsc, T\})$
      \State $\psi \sim p(\psi)$
      \State $\bm{\epsilon} \sim \mathcal{N}(\mathbf{0}, \mathbf{I})$
      \State $\vy \leftarrow $U-Krig$(\tilde{\vy})$
      \State Compute $\Delta_{\bx_t}(\bm{h})$ and $\Delta_{\bx_0}(\bm{h})$
      \State $\mathcal{R}_V = \frac{1}{|\mathcal{H}|} \sum_{\bm{h} \in \mathcal{H}} (\Delta_{\bx_t}(\bm{h}) - \Delta_{\bx_0}(\bm{h}))^2$
      \State Take a gradient descent step on
      \Statex $\grad_\theta (\left\| \bepsilon - f_\theta(\sqrt{\psi_t} \bx_0 + \sqrt{1-\psi_t} \bepsilon, \vy, t) \right\|^2 + \lambda_V \mathcal{R}_V)$
    \Until{converged}
  \end{algorithmic}
\end{algorithm}
\end{minipage}
\Description{An algorithm describing the training process in the Kriging-informed Conditional Diffusion Probabilistic Model (Ki-CDPM), including steps such as sampling, applying U-Krig, computing variogram regularization, and updating model parameters via gradient descent.}
\vspace*{-.2cm}
\end{figure}

Once the Kriging-informed Conditional Diffusion Probabilistic Model (Ki-CDPM) has been trained, it can be used for inference to generate high-resolution sea-level elevation maps from coarse-resolution inputs. The inference involves applying the learned reverse diffusion process to a given coarse-resolution map (as a conditional input) to obtain a detailed, spatially coherent high-resolution map. Algorithm \ref{alg:sampling2} describes the steps involved in the inference process and discusses generating high-resolution sea-level elevation maps using the Ki-CDPM.

\begin{algorithm}[H]
\caption{Inference in Ki-CDPM} \label{alg:sampling2}
\begin{algorithmic}[1]
% \State Sample a coarse-resolution elevation map $Y \sim q(Y)$
% \State Estimate variogram parameters $\theta_{UK}$ from $Y$ using MLE
% \State Perform U-Krig on $Y$ with $\theta_{UK}$ to obtain $\tilde{X} \in \mathbb{R}^{N \times N}$
\State $\vx_T \sim \mathcal{N}(\mathbf{0}, \mathbf{I})$
\State $\vy \leftarrow $U-Krig$(\tilde{\vy})$
\For{$t=T, \dotsc, 1$}
\State $\vz \sim \mathcal{N}(\mathbf{0}, \mathbf{I})$ if $t > 1$, else $\vz = \mathbf{0}$
\State $\vx_{t-1} = \frac{1}{\sqrt{\alpha_t}}\left( \vx_t - \frac{1-\alpha_t}{\sqrt{1-\psi_t}} f_\theta(\vx_t, \vy, \psi_t) \right) + \sqrt{1 - \alpha_t} \vz$
\EndFor
\State \textbf{return} $\vx_{0}$
\end{algorithmic}
\end{algorithm}

% The inference process leverages the trained Ki-CDPM to generate a high-resolution sea-level elevation map consistent with the coarse-resolution input. It exhibits the spatial dependence structure informed by the estimated variogram parameters. The Kriging initialization and the parametric map construction steps ensure that the generated map is spatially coherent and captures the large-scale patterns present in the coarse-resolution data.

% This section presents a detailed evaluation of the Kriging-informed Conditional Diffusion Probabilistic Model (Ki-CDPM) for downscaling climate variable (sea-level elevation) data from a coarse resolution of 1 degree to a fine resolution of 0.25 degrees. We compare the performance of Ki-CDPM with several baseline methods, including Bicubic Interpolation, CNN-based downscaling \cite{vandal2017deepsd}, GAN-based downscaling \cite{harris2022generative}, and Diffusion-based downscaling \cite{watt2024generative}. The evaluation metrics used for comparison include Root Mean Square Error (RMSE), Mean Absolute Error (MAE), Pearson Correlation Coefficient (PCC), and Continuous Ranked Probability Score (CRPS).

\section{Experimental Evaluation}
\label{sec:ExperimentalEvaluation}
\textbf{Experimental Goal:} Our experimental goal was to compare the solution quality of downscaling from our proposed Ki-CDPM model against state-of-the-art downscaling methods and provide both qualitative and quantitative analysis.
\subsection{Experiment Design}
% \textcolor{red}{[Aneesh could you please add a few lines on why we are looking at specific regions for the sea-level elevation and not the whole US or downscale globally]}
\textbf{Datasets:} Our experimental evaluation focused on downscaling two key climate variables: sea-level anomaly (SLA) and eddy kinetic energy (EKE). We use high-resolution Copernicus and CMIP6 datasets and examined various sub-regions, including Eastern North America (ENA), western North America (WNA), and the Bay of Bengal (BoB)  \cite{Iturbide2020}, an area particularly vulnerable to coastal flooding, due to the absence of ground truth data for climate models and the need for bias correction. We tested our methodology on satellite observations where the ground truth is known. Climate model outputs on sea level change are biased in the mean and variability, and the ground truth is unknown for high-resolution sea level values. Hence, we did not downscale climate model output in this study. Future studies will utilize high-resolution climate model outputs for training datasets as these high-resolution models are currently being tested\cite{Gascon2023}.

The Copernicus Climate Data Store (CDS) dataset offers comprehensive global sea level anomaly data derived from satellite altimetry measurements. This dataset spans from 1993 to now and provides daily and monthly mean estimates of sea level anomalies. These anomalies are calculated with respect to a twenty-year reference sea level using absolute standards. The data is essential for monitoring the long-term evolution of sea levels and analyzing ocean and climate indicators. It includes sea level anomalies, absolute dynamic topography, and geostrophic velocities, which are crucial for approximating ocean surface currents. The dataset is updated approximately three times a year with a delay of about five months to ensure accuracy and stability \cite{CDS_portfolio, C3S_portfolio}.

The CMIP6 HighResMIP versions of EC-Earth provide global high-resolution coupled climate data developed by the EC-Earth consortium. The dataset includes EC-Earth3P-HR, with a high resolution of approximately 40 km for the atmosphere and 0.25 degrees for the ocean, and a standard-resolution version, EC-Earth3P, with 80 km for the atmosphere and 1.0 degrees for the sea. These are part of the High-Resolution Model Intercomparison Project (HighResMIP) and are designed to improve the accuracy of climate simulations by using higher resolutions. The high resolution enhances the representation of certain climate phenomena like the El Niño–Southern Oscillation, although it does not universally reduce biases in all regions \cite{Haarsma2020}.

\begin{figure}[ht]
    \centering
    \includegraphics[width=0.48\textwidth]{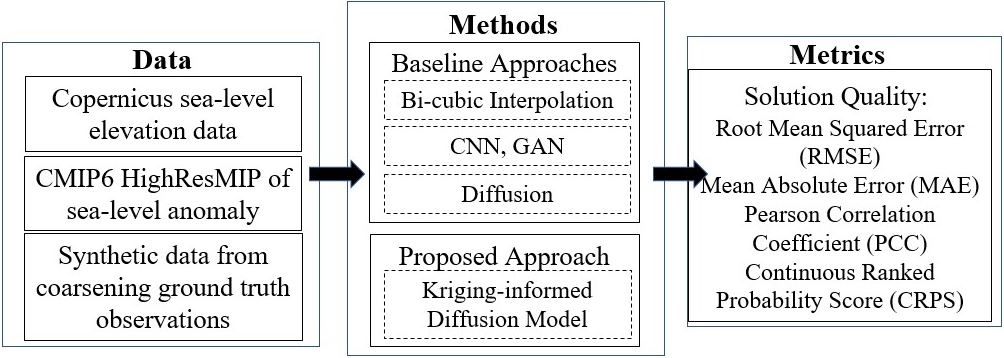}
    \caption{Experiment Design}
    \Description{A diagram illustrating the experimental design, showing the flow and structure of the experimental setup.}
    \label{fig:ExperimentDesign}
\end{figure}

% \subsubsection{Baselines:} Recent advancements in climate variable downscaling have predominantly leveraged deep learning techniques, with diffusion models emerging as the current state-of-the-art approach.
% \begin{itemize}
%     \item \textbf{Bicubic Interpolation:} A standard image interpolation method that uses cubic convolution to estimate pixel values.
%     \item \textbf{CNN-based Downscaling\cite{vandal2017deepsd}:} A CNN model trained to enhance the resolution of low-resolution images using deconvolution.
%     \item \textbf{GAN-based Downscaling\cite{harris2022generative}:} A GAN approach where a generator produces high-resolution images and a discriminator evaluates their quality.
%     \item \textbf{Diffusion-based Downscaling\cite{watt2024generative}:} A baseline diffusion model similar to Denoising Diffusion Probabilistic Models (DDPM) for generating high-resolution images.
% \end{itemize}

\textbf{Training \& Inference:} The model was trained on monthly mean data for both climate variables (sla, eke) in each of the three regions (ENA, WNA, BoB) for the duration from 1993-02 to 2013-12. The range of years for evaluation spans from 2014-01 to 2023-05. The data was transformed to the range [-1,1] to facilitate training convergence. We fixed \( T = 1000 \) for all experiments to align the number of neural network evaluations during sampling with those in prior studies~\cite{ho2020, sohl2015deep, song2019generative}. The variances in the forward process were set to constants that increase linearly from \( \beta_1 = 10^{-4} \) to \( \beta_T = 0.02 \). These values were selected to be small in comparison to the data scaled to the range \([-1, 1]\), ensuring that the reverse and forward processes have roughly the same functional form while maintaining the signal-to-noise ratio at \( \mathbf{x}_T \) as low as possible.

\textbf{Evaluation Metrics:} We assess model performance using standard downscaling evaluation metrics: root mean square error (RMSE), mean absolute error (MAE), and Pearson correlation coefficient (PCC). Additionally, we employed the continuous ranked probability score (CRPS) to evaluate the uncertainty inherent in the multiple predictions generated by diffusion models due to sampling from the terminal Gaussian distribution \( \mathbf{x}_T \). As only the traditional diffusion model in our baseline models produces probabilistic predictions, CRPS was used exclusively for comparison with this model.

% \begin{itemize}
%     \item \textbf{Root Mean Square Error (RMSE)}
%     % \begin{equation}
%     % \text{RMSE} = \sqrt{\frac{1}{N} \sum_{i=1}^{N} (\hat{y}_i - y_i)^2}
%     % \end{equation}
%     % RMSE measures the square root of the average squared differences between predicted and observed values.

%     \item \textbf{Mean Absolute Error (MAE)}
%     % \begin{equation}
%     % \text{MAE} = \frac{1}{N} \sum_{i=1}^{N} |\hat{y}_i - y_i|
%     % \end{equation}
%     % MAE measures the average absolute differences between predicted and observed values.

%     \item \textbf{Pearson Correlation Coefficient (PCC):} PCC measures the linear relationship between predicted and observed values.
%     % \begin{equation}
%     % \text{PCC} = \frac{\sum_{i=1}^{N} (\hat{y}_i - \bar{\hat{y}}) (y_i - \bar{y})}{\sqrt{\sum_{i=1}^{N} (\hat{y}_i - \bar{\hat{y}})^2 \sum_{i=1}^{N} (y_i - \bar{y})^2}}
%     % \end{equation}

\textbf{Continuous Ranked Probability Score (CRPS):} 
    \begin{equation}
    \text{CRPS}(F, x) = \int_{-\infty}^{\infty} \left( F(y) - \mathbf{1}\{y \geq x\} \right)^2 dy
    \end{equation}
    where \( \mathbf{1}\{\cdot\} \) is the indicator function.
    CRPS \cite{hersbach2000decomposition} is employed to evaluate the accuracy and reliability of probabilistic forecasts in our downscaling model. CRPS measures the difference between the cumulative distribution function (CDF) of the predicted probability distribution \( F \) and the CDF of the observed value \( x \). A lower CRPS value indicates a forecast that closely matches the observed outcomes, effectively capturing the uncertainty and variability in the predictions. This metric is beneficial since it provides a comprehensive measure of forecast quality, encompassing both the accuracy and sharpness of the probabilistic predictions. Figure \ref{fig:ExperimentDesign} shows the overall experiment design. 
% \end{itemize}

% \begin{table}[h]
%     \centering
%     % \footnotesize % Further reduce the font size
%     \begin{tabular}{p{3.8cm}cccc}
%         \toprule
%         \textbf{Method} & \textbf{RMSE$\downarrow$} & \textbf{MAE$\downarrow$} & \textbf{PCC$\uparrow$} & \textbf{CRPS$\downarrow$} \\
%         \midrule
%         Bicubic Interpolation & 0.87 & 0.68 & 0.72 & - \\
%         CNN-based Downscaling & 0.43 & 0.35 & 0.81 & - \\
%         GAN-based Downscaling & 0.28 & 0.24 & 0.87 & - \\
%         % U-Net based Downscaling & 0.027 & 0.019 & 0.91 \\
%         Diffusion-based Downscaling & 0.15 & 0.11 & 0.91 & 0.12 \\
%         \textbf{Ki-CDPM} & \textbf{0.11} & \textbf{0.08} & \textbf{0.94} & \textbf{0.09} \\
%         \bottomrule
%     \end{tabular}
%     \caption{Performance comparison of Ki-CDPM and baseline methods using RMSE, MAE, PCC, and CRPS.}
%     \label{tab:performance}
% \end{table}
\subsection{Experimental Results}

\begin{table*}[!ht]
\centering
\caption{Performance comparison of Ki-CDPM and baseline methods using RMSE, MAE, and PCC on sea-level elevation.}
\begin{tabular}{lccccccccccc}
\toprule
\multirow{2}{*}{\textbf{Model}} & \multicolumn{3}{c}{Eastern North America (ENA)} & & \multicolumn{3}{c}{Western North America (WNA)} & & \multicolumn{3}{c}{Bay of Bengal (BoB)} \\
\cmidrule{2-4} \cmidrule{6-8} \cmidrule{10-12}
 & \textbf{RMSE$\downarrow$} & \textbf{MAE$\downarrow$} & \textbf{PCC$\uparrow$} & & \textbf{RMSE$\downarrow$} & \textbf{MAE$\downarrow$} & \textbf{PCC$\uparrow$} & & \textbf{RMSE$\downarrow$} & \textbf{MAE$\downarrow$} & \textbf{PCC$\uparrow$} \\
\midrule
Bicubic Interpolation & 18.7 & 14.8 & 0.72 & & 25.1 & 18.3 & 0.68 & & 29.7 & 22.9 & 0.65 \\
CNN-based Downscaling \cite{vandal2017deepsd} & 6.3 & 5.6 & 0.81 & & 11.4 & 8.8 & 0.78 & & 13.4 & 11.8 & 0.74 \\
GAN-based Downscaling \cite{harris2022generative} & 4.8 & 4.1 & 0.87 & & 8.3 & 6.2 & 0.81 & & 9.8 & 8.3 & 0.78 \\
Baseline Diffusion Downscaling \cite{watt2024generative} & 2.5 & 1.9 & 0.91 & & 4.6 & 3.4 & 0.87 & & 6.1 & 5.1 & 0.82 \\
\textbf{Ki-CDPM} & \textbf{1.1} & \textbf{0.8} & \textbf{0.94} & & \textbf{2.4} & \textbf{2.1} & \textbf{0.91} & & \textbf{4.0} & \textbf{3.2} & \textbf{0.86} \\
\bottomrule
\end{tabular}

\label{tab:performance_comparison_sla}
\end{table*}

\begin{table}[!ht]
\centering
\caption{Performance comparison of Ki-CDPM and baseline methods using CRPS on sea-level elevation
.}
\begin{tabular}{lccc}
\toprule
\multirow{2}{*}{\textbf{Model}} & \multicolumn{3}{c}{CRPS$\downarrow$} \\
\cmidrule{2-4}
 & \textbf{ENA} & \textbf{WNA} & \textbf{BoB} \\
\midrule
Baseline Diffusion Downscaling & 0.13 & 0.31 & 0.37 \\
\textbf{Ki-CDPM} & \textbf{0.09} & \textbf{0.22} & \textbf{0.31} \\
\bottomrule
\end{tabular}
\label{tab:performance_comparison_CRPS_sla}
\end{table}

\begin{table}[!ht]
    \centering
    \caption{Ablation study on the impact of variogram regularizer in Ki-CDPM on sea-level elevation in Eastern North America (ENA).}
    % \small % Reduce font size
    \begin{tabular}{p{3.35cm}cccc}
        \hline
        \textbf{Method} & \textbf{RMSE$\downarrow$} & \textbf{MAE$\downarrow$} & \textbf{PCC$\uparrow$} & \textbf{CRPS$\downarrow$} \\
        \hline
        \textbf{Ki-CDPM (w variogram)} & \textbf{1.1} & \textbf{0.8} & \textbf{0.94} & \textbf{0.09} \\
        Ki-CDPM (w/o variogram) & 1.3 & 1.0 & 0.92 & 0.11 \\
        \hline
    \end{tabular}
    \label{tab:ablation_study_sla}
\end{table}

\begin{table*}[!ht]
\centering
\caption{Performance comparison of Ki-CDPM and baseline methods using RMSE, MAE, and PCC on eddy kinetic energy (EKE).}
\begin{tabular}{lccccccccccc}
\toprule
\multirow{2}{*}{\textbf{Model}} & \multicolumn{3}{c}{Eastern North America (ENA)} & & \multicolumn{3}{c}{Western North America (WNA)} & & \multicolumn{3}{c}{Bay of Bengal (BoB)} \\
\cmidrule{2-4} \cmidrule{6-8} \cmidrule{10-12}
 & \textbf{RMSE$\downarrow$} & \textbf{MAE$\downarrow$} & \textbf{PCC$\uparrow$} & & \textbf{RMSE$\downarrow$} & \textbf{MAE$\downarrow$} & \textbf{PCC$\uparrow$} & & \textbf{RMSE$\downarrow$} & \textbf{MAE$\downarrow$} & \textbf{PCC$\uparrow$} \\
\midrule
Bicubic Interpolation & 1524.67 & 1207.52 & 0.71 & & 1762.34 & 1415.88 & 0.67 & & 1985.19 & 1603.27 & 0.64 \\
CNN-based Downscaling & 752.14 & 621.87 & 0.82 & & 869.42 & 733.64 & 0.77 & & 1052.76 & 895.41 & 0.74 \\
GAN-based Downscaling & 493.58 & 425.36 & 0.86 & & 647.25 & 557.43 & 0.82 & & 751.09 & 663.71 & 0.79 \\
Baseline Diffusion Downscaling & 266.27 & 219.84 & 0.92 & & 411.92 & 347.66 & 0.88 & & 486.91 & 419.52 & 0.83 \\
\textbf{Ki-CDPM} & \textbf{194.83} & \textbf{158.62} & \textbf{0.95} & & \textbf{293.51} & \textbf{246.83} & \textbf{0.91} & & \textbf{354.07} & \textbf{301.14} & \textbf{0.86} \\
\bottomrule
\end{tabular}
\label{tab:performance_comparison_eke}
\end{table*}

\begin{table}[!ht]
\centering
\caption{Performance comparison of Ki-CDPM and baseline methods using CRPS on eddy kinetic energy (EKE).}
\begin{tabular}{lccc}
\toprule
\multirow{2}{*}{\textbf{Model}} & \multicolumn{3}{c}{CRPS$\downarrow$} \\
\cmidrule{2-4}
 & \textbf{ENA} & \textbf{WNA} & \textbf{BoB} \\
\midrule
Baseline Diffusion Downscaling & 0.14 & 0.30 & 0.39 \\
\textbf{Ki-CDPM} & \textbf{0.08} & \textbf{0.23} & \textbf{0.30} \\
\bottomrule
\end{tabular}
\label{tab:performance_comparison_CRPS_eke}
\end{table}
\begin{table}[!ht]
    \centering
    \caption{Ablation study on the impact of variogram regularizer in Ki-CDPM on eddy kinetic energy (EKE) in Eastern North America (ENA).}
    % \small % Reduce font size
    \begin{tabular}{p{3.4cm}cccc}
        \hline
        \textbf{Method} & \textbf{RMSE$\downarrow$} & \textbf{MAE$\downarrow$} & \textbf{PCC$\uparrow$} & \textbf{CRPS$\downarrow$} \\
        \hline
        \textbf{Ki-CDPM w/ variogram)} & \textbf{194.83} & \textbf{158.62} & \textbf{0.95} & \textbf{0.08} \\
        Ki-CDPM (w/o variogram) & 250.25 & 198.27 & 0.93 & 0.11 \\
        \hline
    \end{tabular}
    \label{tab:ablation_study_eke}
\end{table}
\textbf{Quantitative results:}
Results averaged over the entire test dataset for Sea-level Anomaly (SLA) in the Eastern North America (ENA) region are presented in Tables \ref{tab:performance_comparison_sla}, \ref{tab:performance_comparison_CRPS_sla}, and \ref{tab:ablation_study_sla}. The quantitative evaluation of the Ki-CDPM model demonstrates its superior performance compared to state-of-the-art baseline methods for sea-level elevation downscaling. Table~\ref{tab:performance_comparison_sla} presents a comprehensive comparison using RMSE, MAE, and PCC metrics across three regions: Eastern North America (ENA), Western North America (WNA), and the Bay of Bengal (BoB). Ki-CDPM consistently achieves the lowest RMSE and MAE values and the highest PCC values among all methods, indicating its ability to generate accurate and spatially consistent downscaled data. The lower RMSE and MAE values indicate higher accuracy in predicting sea-level elevation. In contrast, the elevated PCC value further suggests that the predicted elevations closely align with the patterns and trends observed in the ground truth data, highlighting the model's ability to accurately capture the spatial variability and gradients inherent in sea-level elevations.

Specifically, in the ENA region, Ki-CDPM obtains an RMSE of 1.1, MAE of 0.8, and PCC of 0.94, outperforming the best-performing baseline, the diffusion-based downscaling method, which yields an RMSE of 2.5, MAE of 1.9, and PCC of 0.91. Similar trends are observed in the WNA and BoB regions, where Ki-CDPM maintains its superiority across all metrics. Table~\ref{tab:performance_comparison_CRPS_sla} compares the performance of Ki-CDPM with the diffusion-based downscaling method using the CRPS metric, which assesses the accuracy and reliability of probabilistic predictions. Ki-CDPM achieves lower CRPS values in all three regions (0.09 for ENA, 0.22 for WNA, and 0.31 for BoB) compared to the diffusion-based method (0.13 for ENA, 0.31 for WNA, and 0.37 for BoB), further confirming its ability to provide more accurate and reliable probabilistic downscaled data.

Additionally, an ablation study presented in Table~\ref{tab:ablation_study_sla} highlights the impact of the variogram regularizer in Ki-CDPM. The inclusion of the regularizer leads to improved performance across all metrics in the ENA region, with an RMSE of 1.1, MAE of 0.8, PCC of 0.94, and CRPS of 0.09, compared to the model without the regularizer (RMSE of 1.3, MAE of 1.0, PCC of 0.92, and CRPS of 0.11). These results underscore the importance of the variogram-based regularizer in enhancing the accuracy and reliability of the Ki-CDPM model for sea-level elevation downscaling.

In addition to downscaling sea-level elevation, the Ki-CDPM model demonstrates superior performance in downscaling other climate variables, such as eddy kinetic energy (EKE), as shown in Tables \ref{tab:performance_comparison_eke}, \ref{tab:performance_comparison_CRPS_eke}, and \ref{tab:ablation_study_eke}. Ki-CDPM achieves the lowest RMSE, MAE, and CRPS values and the highest PCC values compared to baseline methods across Eastern North America (ENA), Western North America (WNA), and Bay of Bengal (BoB) regions. Specifically, in the ENA region, Ki-CDPM with variogram regularizer achieves an RMSE of 194.83, MAE of 158.62, PCC of 0.95, and CRPS of 0.08, significantly outperforming the model without the regularizer. These results highlight the versatility and robustness of Ki-CDPM in accurately downscaling diverse climate variables.
\begin{figure*}[!t]
\begin{minipage}{0.08\textwidth}
    \centering
    \includegraphics[width=\textwidth,height=7cm]{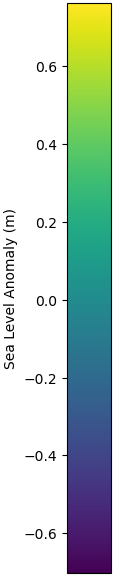}\\
    \vfill
    \vspace{25px}
    \includegraphics[width=\textwidth,height=5cm]{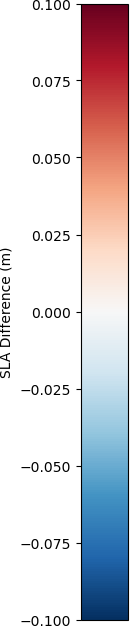}
\end{minipage}
\begin{minipage}{0.9\textwidth}
  \centering
  \begin{subfigure}{0.4\textwidth}
    \includegraphics[width=\textwidth]{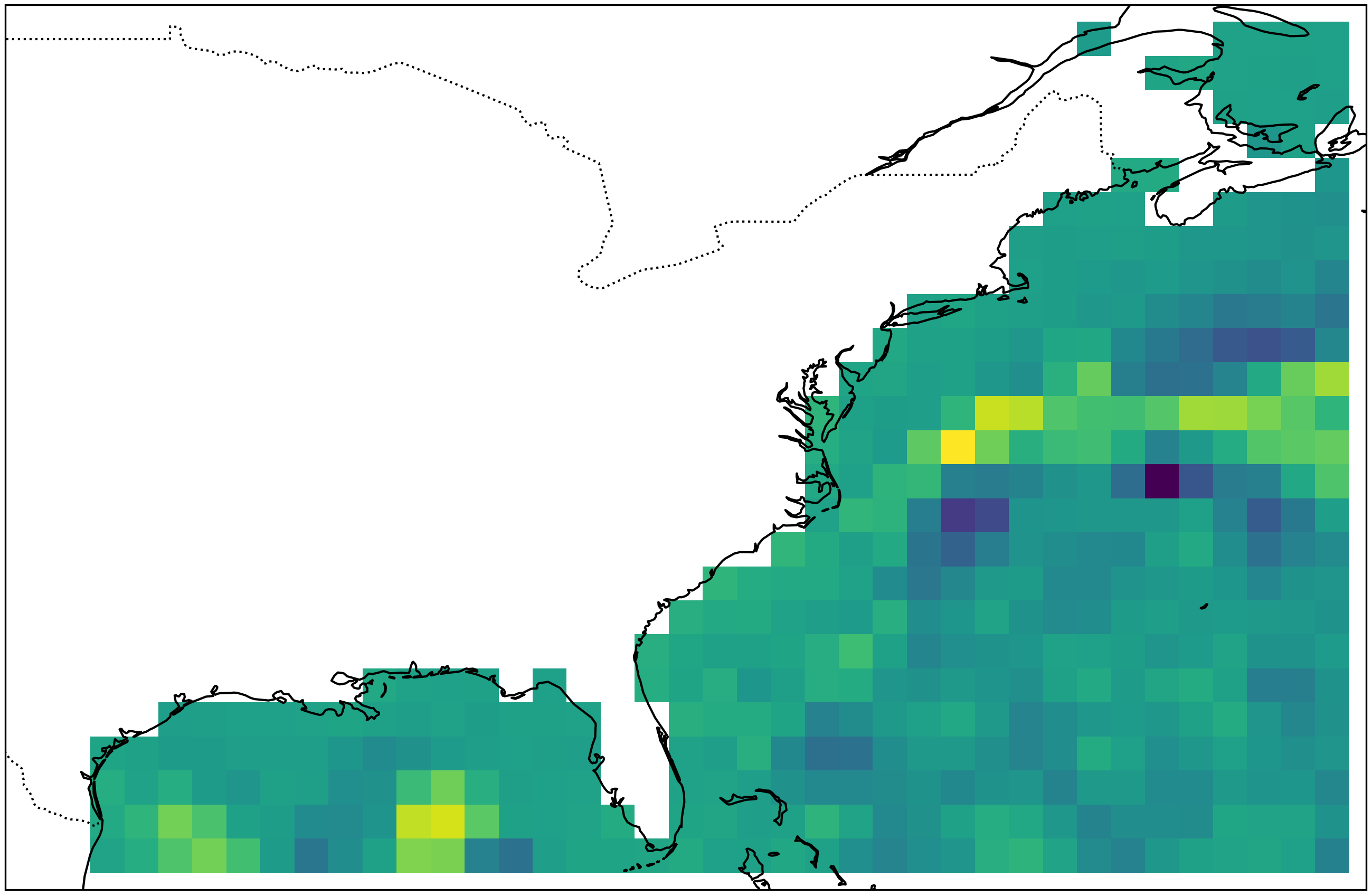}
    \caption{Coarse-Resolution Input Data}
    \label{fig:coarse_sla_image_viz}
    \Description{A visualization of the coarse-resolution sea-level anomaly input data.}
  \end{subfigure}
  \vspace{2px} % Adds vertical space between rows of subfigures
  \begin{subfigure}{0.4\textwidth}
    \includegraphics[width=\textwidth]{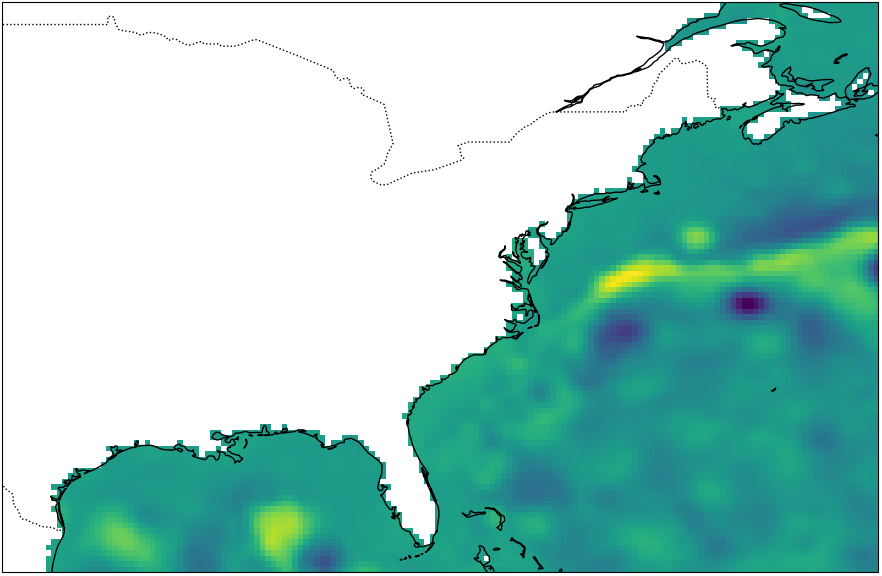}
    \caption{Ground Truth High-Res. Data}
    \label{fig:ground_truth_viz}
    \Description{A visualization of the high-resolution ground truth sea-level anomaly data.}
  \end{subfigure}
  \vspace{2px} % Adds vertical space between rows of subfigures
  \begin{subfigure}{0.4\textwidth}
    \includegraphics[width=\textwidth]{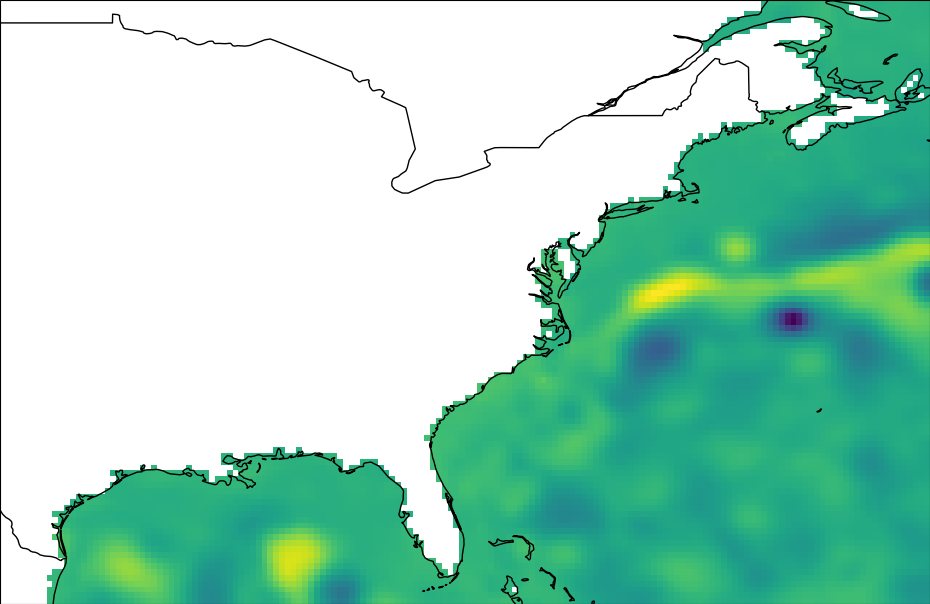}
    \caption{Baseline Diffusion Output}
    \label{fig:Diffusion_output_viz}
    \Description{A visualization showing the output from the baseline diffusion model for sea-level anomaly.}
  \end{subfigure}
  \begin{subfigure}{0.4\textwidth}
    \includegraphics[width=\textwidth]{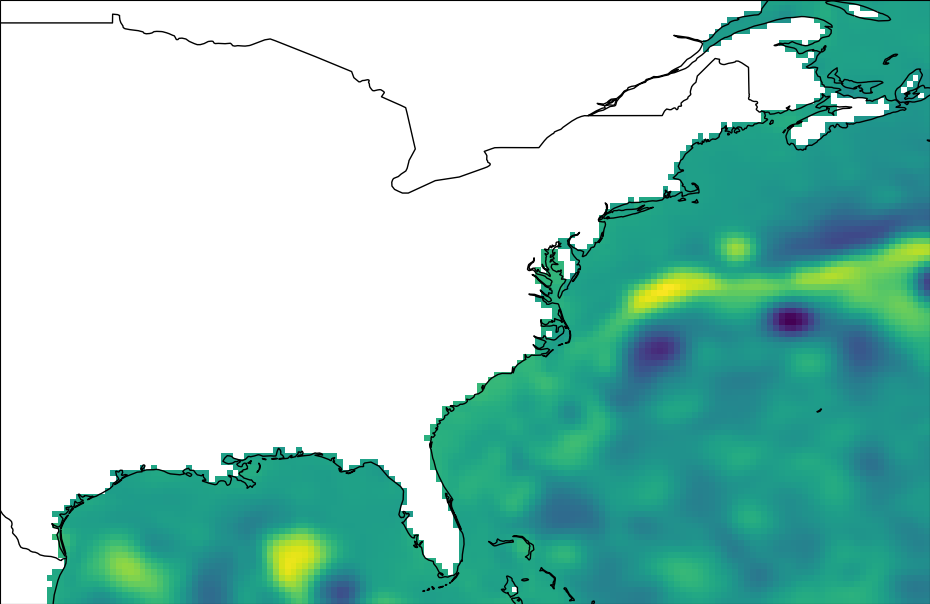}
    \caption{Ki-CDPM Output}
    \label{fig:Ki-CDPM_output_viz}
    \Description{A visualization of the output from the Kriging-informed Conditional Diffusion Probabilistic Model (Ki-CDPM) for the sea-level anomaly.}
  \end{subfigure}
  \begin{subfigure}{0.4\textwidth}
    \includegraphics[width=\textwidth]{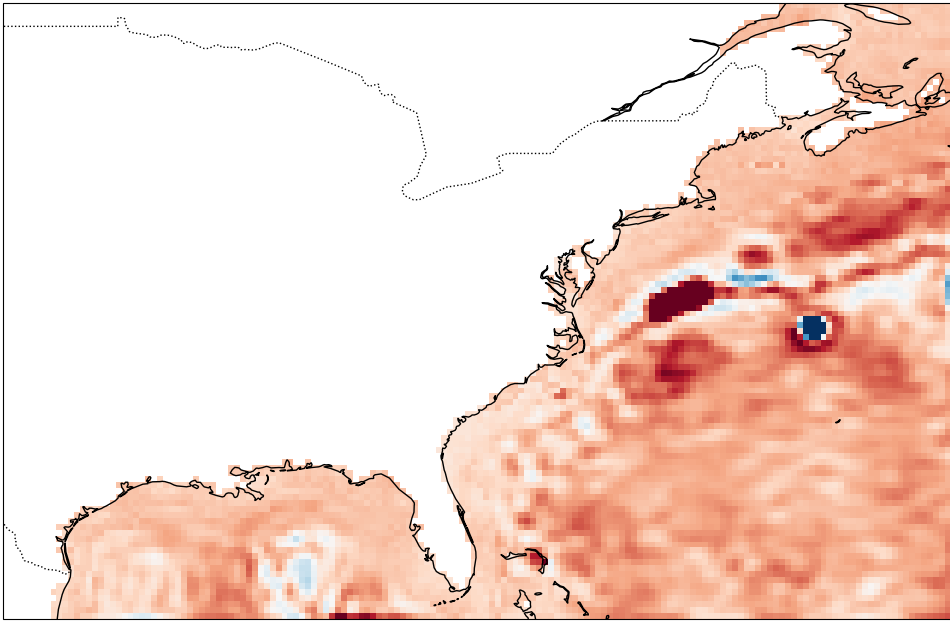}
    \caption{Difference b/w baseline diffusion \& ground truth}
    \label{fig:GT_TD_sla}
    \Description{A difference map between the baseline diffusion output and the ground truth for sea-level anomaly.}
  \end{subfigure}
  \begin{subfigure}{0.4\textwidth}
    \includegraphics[width=\textwidth]{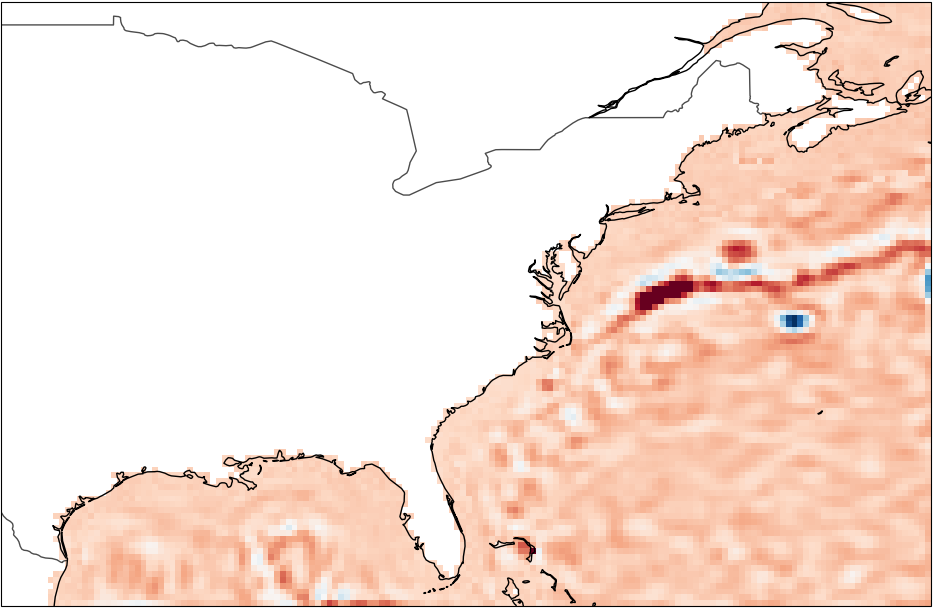}
    \caption{Difference b/w Ki-CDPM \& ground truth}
    \label{fig:extra_image_2}
    \Description{A difference map between the Ki-CDPM output and the ground truth for sea-level anomaly.}
  \end{subfigure}
\end{minipage}

  \caption{Qualitative Analysis on sea-level elevation in ENA region \textit{(Best in color)}.}
  \Description{This figure shows a qualitative analysis of sea-level elevation in the ENA region with six subfigures: coarse-resolution input data, high-resolution ground truth data, baseline diffusion output, Ki-CDPM output, and two difference maps comparing baseline diffusion and Ki-CDPM outputs with the ground truth.}
  \label{fig:Qualitative_Analysis1_sla}
\end{figure*}

\textbf{Qualitative results:} Figure \ref{fig:Qualitative_Analysis1_sla} illustrates the comparative performance of Ki-CDPM against the ground truth and baseline diffusion model for sea-level elevation at a single timestep. The baseline model, conditioned on bicubic interpolation of coarse data, exhibits limitations in predicting values closer to the coastline and produces some pixelated artifacts. In contrast, Ki-CDPM leverages the spatial dependency structure provided by Kriging and the variogram, resulting in superior downscaling with finer details and improved accuracy. Resolving these fine-scale structures and gradients in sea-level elevation is essential to predict better regional impacts of sea-level rise and better model spatial variability in sea-level elevation. The fine-scale eddy features seen in the downscaled maps of sea level elevation play a crucial role in ocean dynamics and the evolution of the ocean fields. The 1 deg coarse resolution model outputs, which is the typical resolution of current generation climate models, do not resolve these features in the ocean. Hence, resolving them and seeing their evolution in time helps predict the changes to ocean circulation and their impact on rising sea levels in coastal communities.

In the study, the satellite observed a high-resolution sea-level elevation map at 0.25 degrees resolution, which serves as the accuracy benchmark, while a coarser 1-degree resolution map is utilized as input data. The main text compares the proposed method's performance with a state-of-the-art traditional diffusion model, showcasing qualitative results.

\begin{figure*}[!t]
  \centering
  \begin{minipage}{0.08\textwidth}
  \centering
    \includegraphics[width=0.75\textwidth,height=7cm]{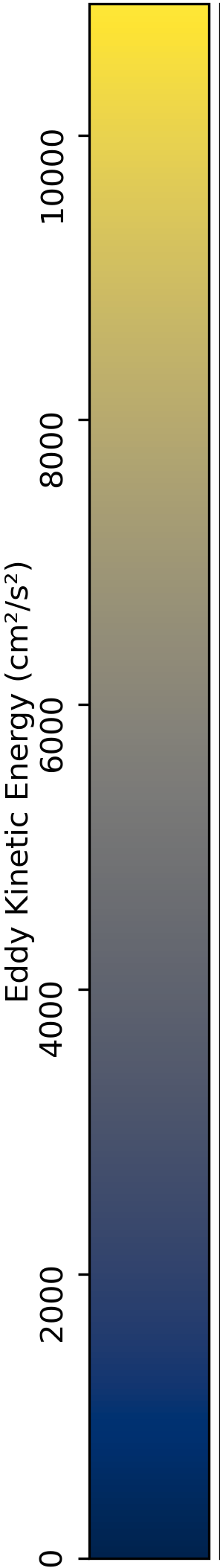}\\
    \vfill
    \vspace{25px}
    \includegraphics[width=0.75\textwidth,height=5cm]{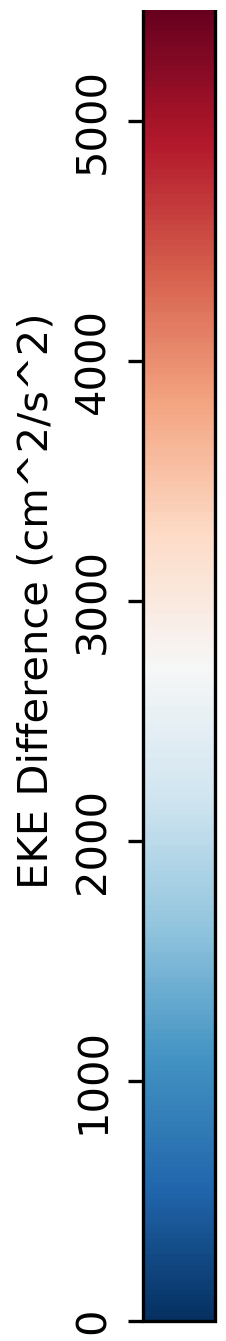}
  \end{minipage}
  \begin{minipage}{0.9\textwidth}
  \centering
  \begin{subfigure}{0.38\textwidth}
    \includegraphics[width=\textwidth]{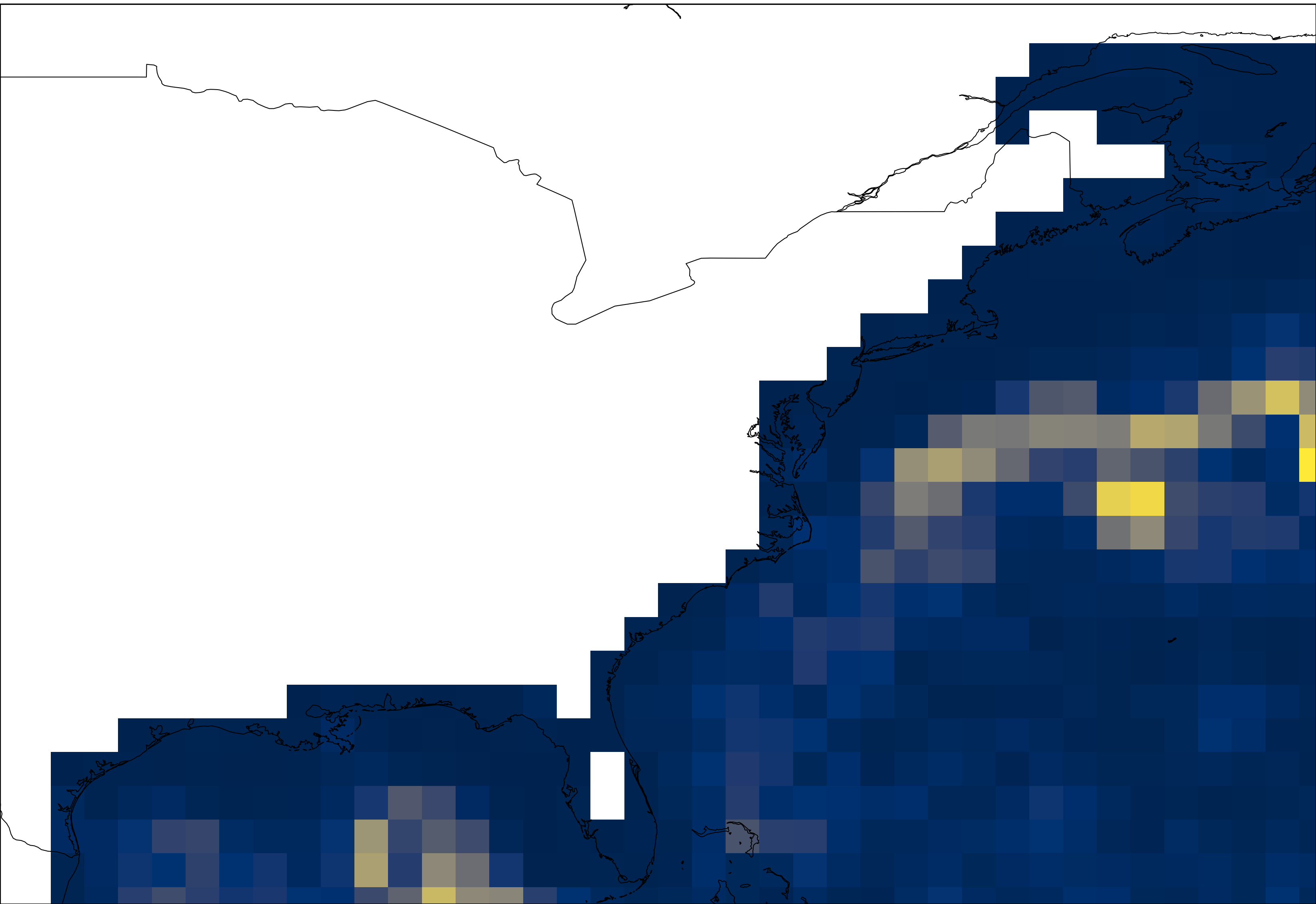}
    \caption{Coarse-Resolution Input Data}
    \label{fig:coarse_eke_image_viz}
    \Description{A visualization of the input data of coarse-resolution eddy kinetic energy (EKE).}
  \end{subfigure}
   \vspace{2px} % Adds vertical space between rows of subfigures
  \begin{subfigure}{0.4\textwidth}
    \includegraphics[width=\textwidth]{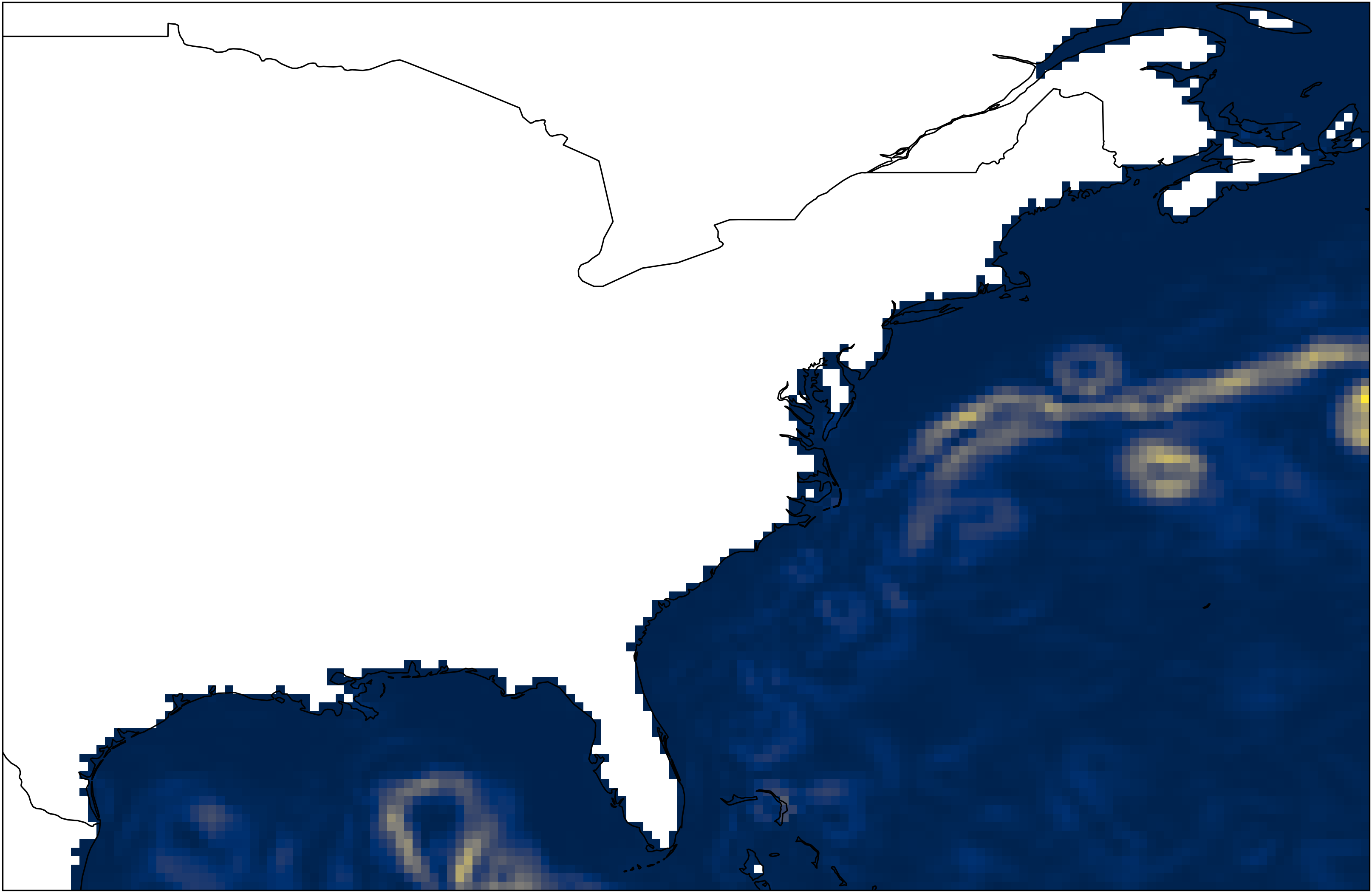}
    \caption{Ground Truth High-Res. Data}
    \label{fig:ground_eke_truth_viz}
    \Description{A visualization of the high-resolution ground truth eddy kinetic energy (EKE) data.}
  \end{subfigure}
   \vspace{2px} % Adds vertical space between rows of subfigures
  \begin{subfigure}{0.4\textwidth}
    \includegraphics[width=\textwidth]{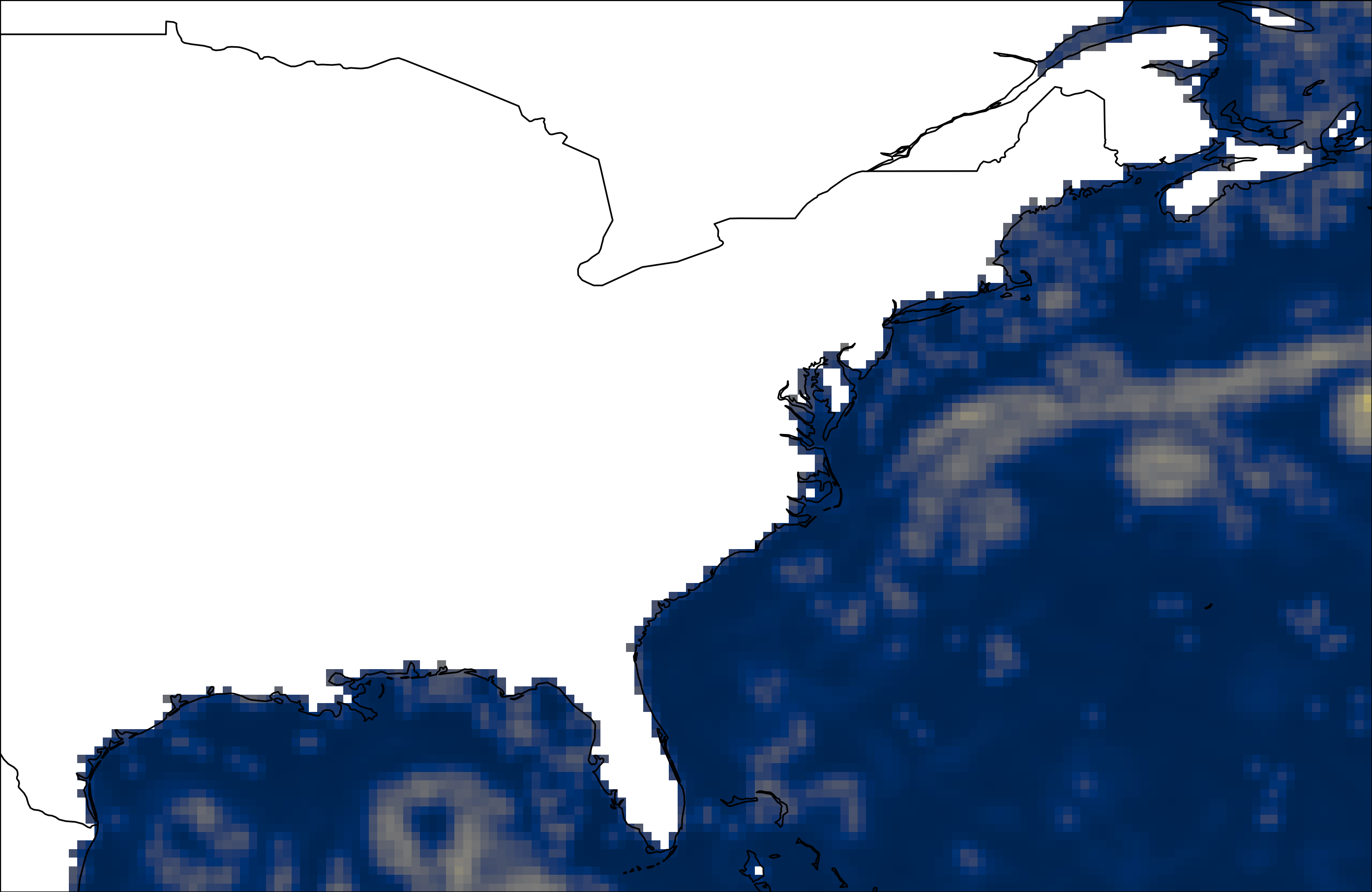}
    \caption{Baseline Diffusion Output}
    \label{fig:Diffusion_eke_output_viz}
    \Description{A visualization showing the output from the baseline diffusion model for eddy kinetic energy (EKE).}
  \end{subfigure}
  \begin{subfigure}{0.4\textwidth}
    \includegraphics[width=\textwidth]{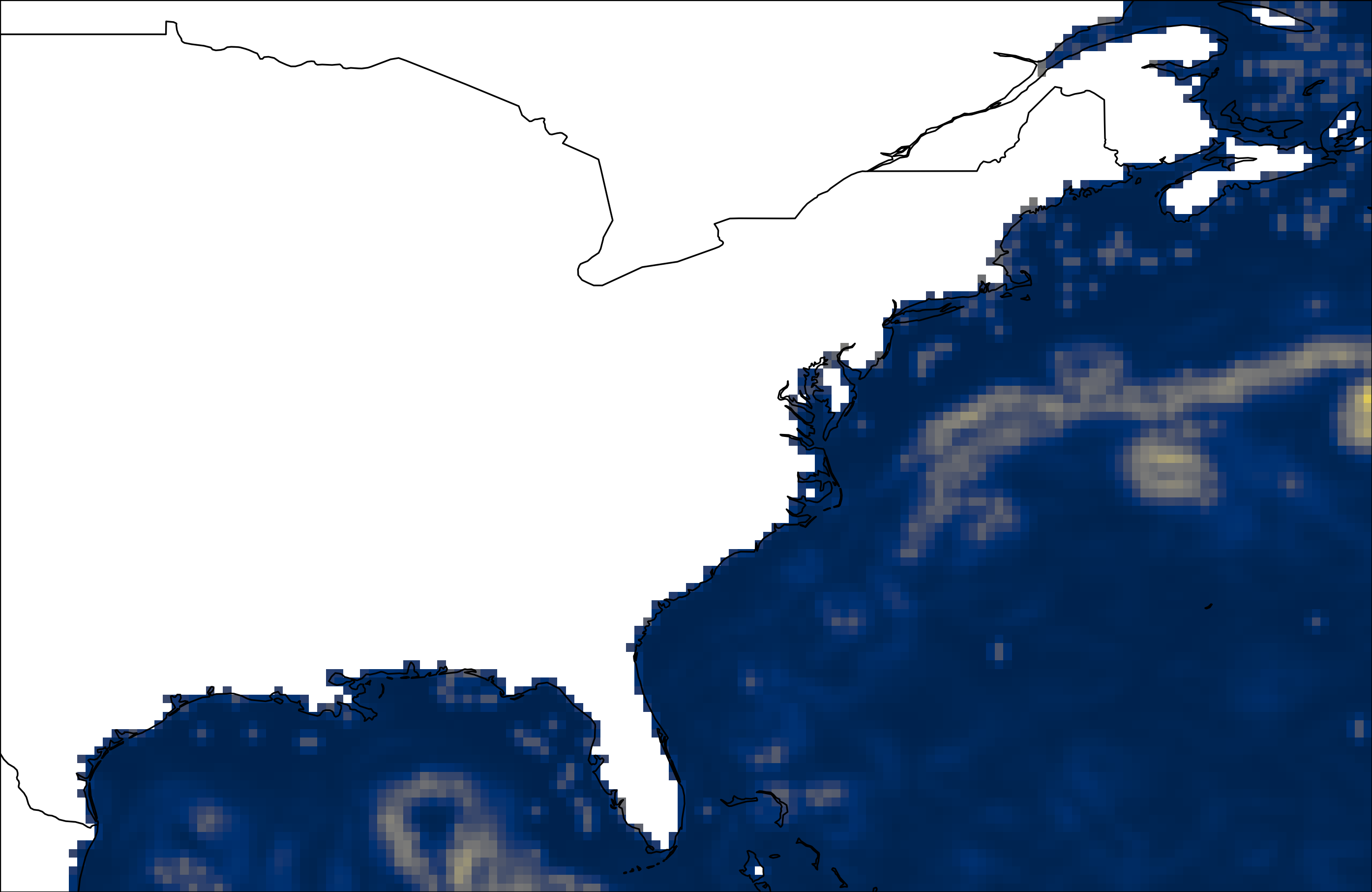}
    \caption{Ki-CDPM Output}
    \label{fig:Ki-CDPM_eke_output_viz}
    \Description{A visualization of the output from the Kriging-informed Conditional Diffusion Probabilistic Model (Ki-CDPM) for eddy kinetic energy (EKE).}
  \end{subfigure}
  
  \begin{subfigure}{0.4\textwidth}
    \includegraphics[width=\textwidth]{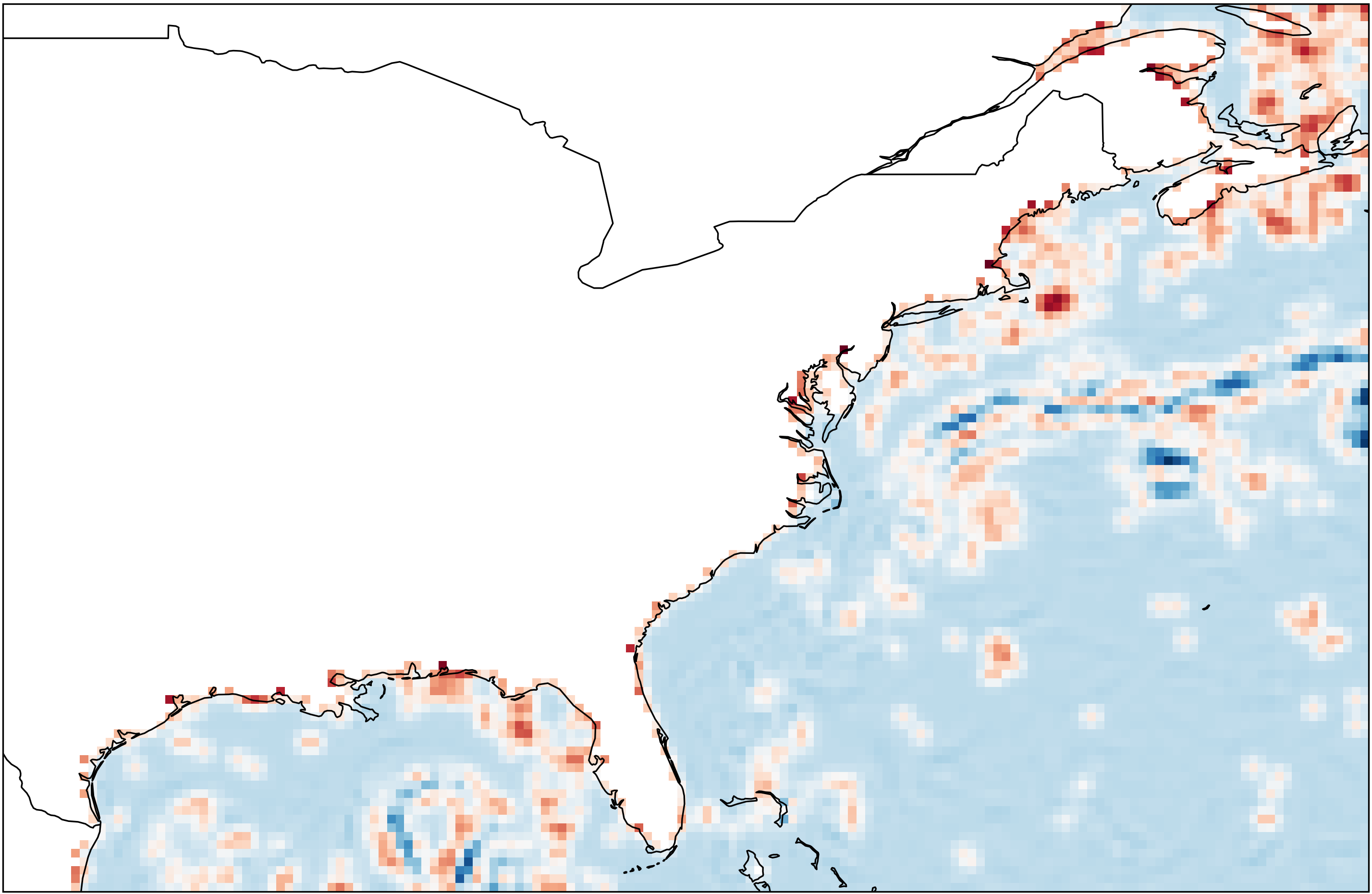}
    \caption{Difference b/w baseline diffusion \& ground truth}
    \label{fig:GT_TD_eke}
    \Description{A difference map between the baseline diffusion output and the ground truth for eddy kinetic energy (EKE).}
  \end{subfigure}
  \begin{subfigure}{0.4\textwidth}
    \includegraphics[width=\textwidth]{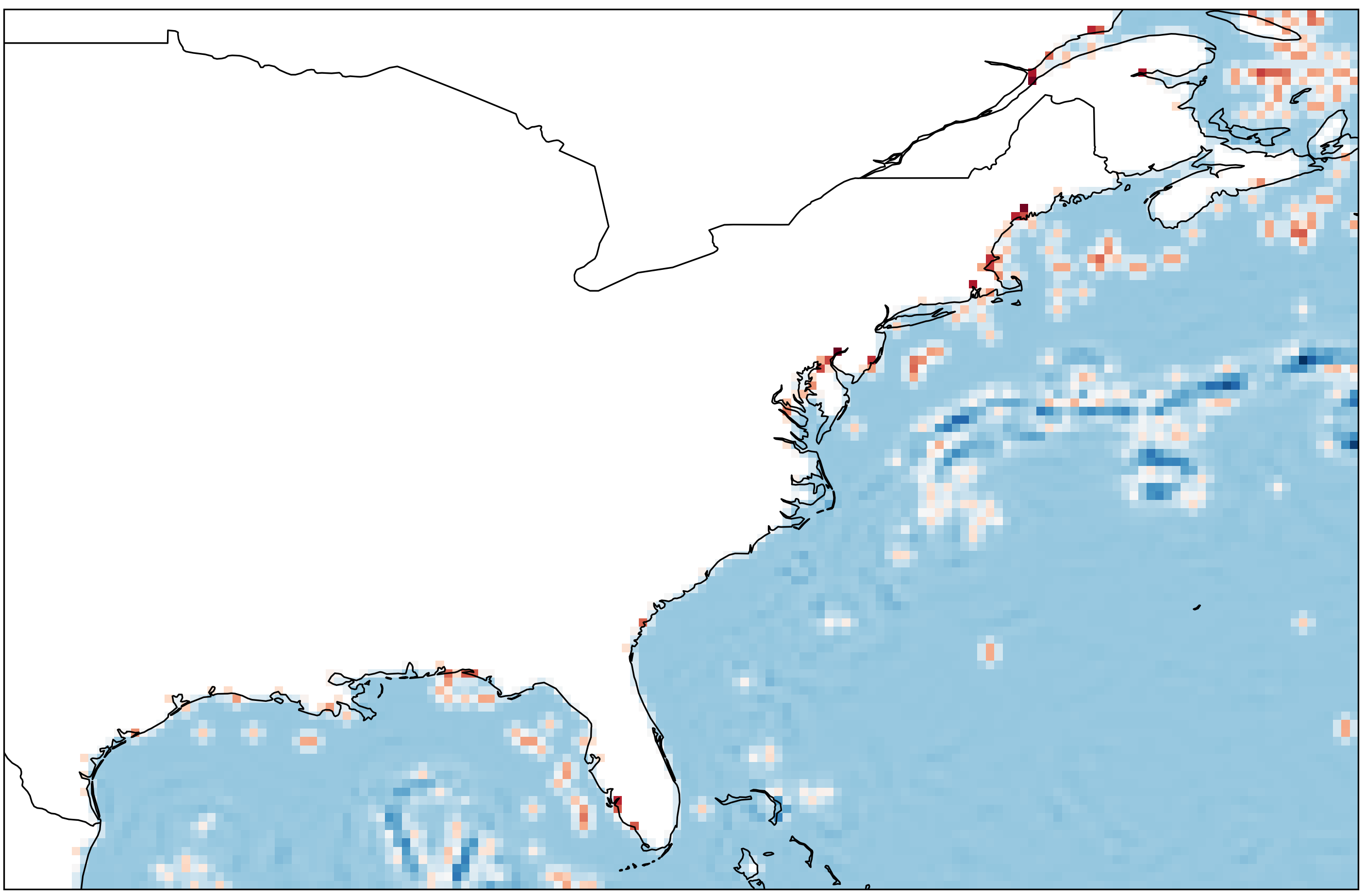}
    \caption{Difference b/w Ki-CDPM \& ground truth}
    \label{fig:GT_kicdpm}
    \Description{A difference map between the Ki-CDPM output and the ground truth for eddy kinetic energy (EKE).}
  \end{subfigure}
\end{minipage}
  \caption{Qualitative Analysis on eddy kinetic energy (EKE) in ENA region \textit{(Best in color)}.}
  \Description{This figure presents a qualitative analysis of eddy kinetic energy (EKE) in the ENA region, including coarse-resolution input data, high-resolution ground truth data, baseline diffusion output, Ki-CDPM output, and two difference maps comparing both models' outputs with the ground truth.}
  \label{fig:Qualitative_Analysis1_eke}
\end{figure*}

% \begin{itemize}
%     \item \textbf{Ground Truth High-Resolution Data:} The actual high-resolution sea-level elevation map (at 0.25 degrees resolution) \cite{https://doi.org/10.24381/cds.4c328c78} is used as the benchmark.
            
%     \item \textbf{Coarse-Resolution Input Data:} The low-resolution sea-level elevation map (at 1-degree resolution) \cite{https://doi.org/10.1029/2019MS001916, https://doi.org/10.24381/cds.4c328c78} is used as the input.
        
%     \item \textbf{Other Baseline Approaches:} In the main paper, we show qualitative results comparing our method with a traditional diffusion model, as it is state of the art in downscaling \cite{watt2024generative}; additional qualitative results on other baseline methods are provided in the appendix.
% \end{itemize}

\section{Related Work}
\label{sec:Related_work}
Spatial variability \cite{gupta2021spatial,farhadloo2024spatial,farhadloo2024towards}, is a significant characteristic of all geographic phenomena, such as climate zones, USDA plant hardiness zones \cite{usda2012usda}, and different terrestrial habitats like forests, grasslands, wetlands, and deserts. This variability influences the flora and fauna within different regions. Additionally, variations in laws, policies, and cultural norms are evident across and within nations. Often referred to as geography's second law, spatial variability is utilized in analytical models like geographically weighted regression (GWR) \cite{mcmillen2004geographically} to measure the interactions between variables in a given area. The challenge of quantifying spatial variability stems from many geophysical elements affecting it. Soil scientists, for example, study soil attributes such as carbon content to evaluate agricultural yield and have noted considerable variation in soil samples within a mere 100 m\textsuperscript{2} due to aspects like tillage, soil makeup, vegetation, land management, and topography. Anomaly detection \cite{chandola2009anomaly} has been widely studied in spatial data mining (e.g., trajectory gaps \cite{sharma2020analyzing, sharma2022towards, sharmaphysics}). However, only a limited number of statistical \cite{ghosh2022towards, ghosh2024towards, ghosh_et_al:LIPIcs.COSIT.2024.25, ghosh_et_al:LIPIcs.GIScience.2023.3}, and machine learning methods have addressed anomaly detection in climate science. 

Traditional machine learning models \cite{mukherjee2019deep}, primarily designed for image processing, encounter several obstacles, including the absence of physical constraints, difficulties with managing high-dimensional climate data, and their inability to yield probabilistic outputs in climate data.
For instance, Autoregressive models (ARs) \cite{oord2016pixel, salimans2017pixelcnn++} are capable of model log-likelihood and complex distributions,  Variational Autoencoders (VAEs) \cite{Kingma2013,rezende2014stochastic} provide rapid sampling capabilities, Generative Adversarial Networks (GANs) \cite{goodfellow2014generative} favored for tasks like class-conditional image generation etc. Climate downscaling \cite{liu2022downscaling, liu2016dynamical, hermans2020improving, kim2021local} is widely used to facilitate understanding and planning for climate impacts at regional or local levels, with machine learning models also applied in statistical downscaling \cite{leinonen2020stochastic}. Recently, diffusion models have gained interest in climate science due to their capacity to model non-linear relationships, although they struggle with generalization, maintaining physical consistency, and non-stationarity. For example, \cite{wan2024debias, watt2024generative} overlook spatial dependencies, leading to inaccurate depictions of physical processes like ocean currents, wind patterns, and temperature gradients. This work introduces Ki-CPDM, incorporating geostatistical capabilities with a Conditional Diffusion Model to effectively capture spatial variability in climate model variables such as sea level rise.

\textbf{Domain Background:} The SWOT (Surface Water and Ocean Topography) \cite{Srinivasan2023} satellite mission offers significant advancements over the TOPEX/Poseidon mission in measuring sea level and surface water elevation. While TOPEX provided accurate ocean surface topography data, SWOT extends this capability to measure ocean and freshwater bodies with unprecedented detail. SWOT's higher-resolution measurements enable precise monitoring of smaller-scale ocean phenomena and inland water bodies, which is crucial for understanding climate change impacts. Its innovative Ka-band Radar Interferometer delivers finer spatial resolution, enhancing our ability to track water storage and movement changes. This comprehensive data set supports improved global water resource management, disaster response, and climate prediction models, addressing modern environmental challenges through advanced computational analysis.

 % For instance, Autoregressive models (ARs) \cite{oord2016pixel, salimans2017pixelcnn++} are capable of model log-likelihood and complex distributions, Normalizing flows \cite{rezende2015variational,dinh2016density, Kingma2018} offer improved sampling speeds and also model exact data likelihood, but their requirement for invertible transformations with computable Jacobian determinants restricts their versatility. Variational Autoencoders (VAEs) \cite{Kingma2013,rezende2014stochastic} provide rapid sampling capabilities, though they generally produce lower image quality compared to GANs and ARs \cite{vahdat2021nvae}. Generative Adversarial Networks (GANs) \cite{goodfellow2014generative} are favored for tasks like class-conditional image generation and super-resolution, but their training often involves complex optimization cycles that require specific techniques for stability \cite{arjovsky-arxiv-2017,gulrajani2017improved}. Additionally, super-resolution tasks typically necessitate a supplementary consistency-based loss to prevent mode collapse \cite{ledig2017photo}. Higher resolution images have been successfully generated using cascaded GAN models \cite{denton-nips-2015}. This work introduces Ki-CPDM, incorporating geostatistical capabilities with a Conditional Diffusion Model to effectively capture spatial variability in climate model variables such as sea level rise.

\section{Conclusion and Future Work}
\label{section:conclusion_future_work}
We proposed the Kriging-Informed Conditional Diffusion Probabilistic Model (Ki-CDPM) to address the challenge of downscaling sea-level elevation data from coarse to fine resolution. We further integrated Universal Kriging via the Mat\'ern variogram model with the Conditional Diffusion Probabilistic Model (CDPM), leveraging the strength of geostatistical interpolation to enhance the resolution and realism of downscaled data. Experimental results demonstrate that Ki-CDPM outperforms state-of-the-art methods, generating high-resolution sea-level projections essential for regional climate impact assessments and coastal management. The official version is published in ACM SIGSPAIIAL 2024\cite{ghosh2024towards}.

\textbf{Future Work:} We will explore the application of the Ki-CDPM to other climate variables, such as temperature and precipitation, to evaluate its versatility and robustness across different datasets. Investigating additional geostatistical models could further improve the quality and diversity of the generated outputs. Moreover, developing strategies incorporating domain knowledge such as ocean bed topology and mass balance will be beneficial. Finally, we plan to optimize the computational efficiency \cite{sharma2018webgiobe,sharma2022towards,farhadloo2025spatiallydelineateddomainadaptedaiclassification} of the diffusion model by using novel sampling methods to reduce its computational demands.

\section{Acknowledgments}
This material is based upon work supported by the National Science Foundation under Grants No. 2118285, approved for public release, 22-536. We also want to thank Kim Koffolt and the spatial computing research group for their helpful comments and refinements.
We also thank NCAR for computing resources.
\bibliographystyle{ACM-Reference-Format}
\bibliography{sample-base}

\appendix

\end{document}